**Streptococcosis in aquaculture: Advances, challenges, and future directions in disease control and prevention**

Hussein Aliu Sule[1,2*], Abdulwakil Olawale Saba[1,3], Choo Yee Yu[4],

[1]*Laboratory of Aquatic Animal Health and Therapeutics, Institute of Bioscience, Universiti Putra Malaysia, 43400 UPM Serdang, Selangor, Malaysia.*
[2]*Department of Biology, Faculty of Science, Confluence University of Science and Technology, 263101 Osara, Nigeria.*
[3]*Department of Fisheries, Faculty of Science, Lagos State University, 102101 Ojo, Lagos State, Nigeria.*
[4]*Laboratory of Vaccine and Biomolecules, Institute of Bioscience, Universiti Putra Malaysia, 43400 UPM Serdang, Selangor, Malaysia.*

*Corresponding author: ibnsulehussein@gmail.com

# Abstract

Aquaculture is pivotal for global food security but faces significant challenges from infectious diseases, particularly those caused by *Streptococcus* species such as *Streptococcus iniae* and *Streptococcus agalactiae*. These pathogens induce severe systemic infections in various fish species, resulting in high morbidity and mortality rates. This review consolidates current knowledge on the epidemiology, pathogenesis, and clinical manifestations of these infections in fish and provides a comprehensive analysis of multifaceted control and prebention strategies. Advancements in genetic engineering and selective breeding are highlighted, demonstrating significant potential in developing disease-resistant fish strains through technologies like CRISPR-Cas9 and genomic selection. We examine the impact of farming practices on disease prevalence, emphasizing the roles of stocking density, feeding regimes, and biosecurity measures. The integration of big data analytics and IoT technologies is shown to revolutionize disease monitoring and management, enabling real-time surveillance and predictive modeling for timely interventions. Progress in vaccine development, including subunit, DNA, and recombinant protein vaccines, highlights the importance of tailored immunoprophylactic strategies. Furthermore, this review emphasizes the One-Health approach and the essential collaboration among industry, academia, and government to address the interconnected health of humans, animals, and the environment. This holistic strategy, supported by advanced technologies and collaborative efforts, promises to enhance the sustainability and productivity of aquaculture systems. Future research directions advocate for continued innovation and interdisciplinary partnerships to overcome the persistent challenges of streptococcal infections in aquaculture.

*Key words:*

Aquaculture, *Streptococcus*, Streptococcosis, GBS, zoonosis, genetic engineering, selective breeding, vaccine development, big data analytics, IoT technologies, One-Health approach, biosecurity measures.

# 1.0   Introduction

The rapid expansion of aquaculture is essential in meeting the rising global demand for seafood and are facing significant challenges from bacterial infections, particularly those caused by *Streptococcus* species. Streptococcosis compromised fish health and productivity, posing severe economic losses to an industry that supplies nearly half of the world's seafood [1]. *Streptococcus* infections, especial]ly from *Streptococcus iniae* and *Streptococcus agalactiae* (Group B Streptococcus, GBS), cause morbidity in various fish species by reducing growth rates, impairing feed conversion ratios, and increasing susceptibility to secondary infections [2,3]. The economic repercussions affect not only from the direct loss of fish due to high mortality rate but also by reducing the overall efficiency and profitability of aquaculture operations.

This review synthesizes findings from a broad range of studies published over the past decade, providing a comprehensive overview of current challenges and advancements in managing *Streptococcus* infections in aquaculture. The scope of the review includes key peer-reviewed articles sourced from major databases such as PubMed, Scopus, and Web of Science, focusing on recent developments in genetic engineering, vaccine development, and disease management strategies. While not exhaustive, the review captures the most significant contributions to the field, offering insights that are both current and relevant.

Despite significant progress, several critical gaps remain unaddressed in the literature. The genetic and molecular mechanisms underlying antibiotic resistance in *Streptococcus* species, for example, are not fully understood, limiting the development of more effective treatments. Additionally, the variability in vaccine efficacy across different contexts, whether due to differences in vaccine formulation, fish species, or environmental conditions, remains unclear. The impact of farming practices on disease prevalence, though recognized, requires further exploration to optimize strategies for disease prevention. Furthermore, while emerging technologies such as genetic engineering, big data analytics, and IoT hold promise for improving disease management, their integration into aquaculture practices is still in its early stages, presenting a significant opportunity for innovation. Finally, although the One-Health approach has been advocated for managing diseases, there is a lack of practical frameworks and case studies demonstrating of its successful implementation in aquaculture settings.

By addressing these gaps, the current review not only consolidates existing knowledge but also proposes innovative solutions that could significantly advance the field. The review offer novel insights and practical recommendations and could be an essential resource for researchers and practitioners alike who aim to enhance the sustainability and productivity of aquaculture systems.

## 2.0   Overview of streptococcosis in aquaculture

### 2.1   Historical Perspective

The history of Streptococcosis in aquaculture has been shaped by significant milestones in disease emergence, scientific advancements, and evolving management strategies. The first documented case of Streptococcosis in fish dates back to 1957, when an outbreak were reported in cultured rainbow trout (*Oncorhynchus mykiss*) in Japan, marking the earliest recognition of *Streptococcus* as an aquatic pathogen [4,5]. Over time, various fish species, including salmon, tilapia, eel, and striped bass, were identified as susceptible to streptococcal infections [5].

In 1972, *Streptococcus iniae* was first isolated from a captive Amazon river dolphin (*Inia geoffrensis*), revealing its potential as a zoonotic pathogen. It was later identified as a major cause of disease in aquaculture, with outbreaks in rainbow trout and tilapia farms in Israel, underscoring the need for enhanced disease control strategies [6]. By the 1980s, *Streptococcus agalactiae* (*Group B Streptococcus*, GBS) emerged as a critical pathogen in tilapia farming, with outbreaks documented across Southeast Asia. This period highlighted the global expansion of streptococcal infections in warm-water aquaculture, exacerbated by antibiotic resistance challenges [5].

The 1990s saw further discoveries, with *Streptococcus ictaluri* emerging as a significant pathogen in North American catfish farming, particularly in the southern United States, where it caused severe economic losses [7]. Around the same time, *Streptococcus parauberis* (formerly *S. uberis* type II) was identified as a pathogen affecting turbot and sea bass in Europe [8]. By 1999, *Streptococcus phocae* was first isolated from Atlantic salmon in Chile, broadening the known impact of *Streptococcus* infections in marine aquaculture [9]. A summary of these historical milestones and their impact on aquaculture management is presented in Table 1.

The zoonotic potential of *S. iniae* was evident in the mid-1990s, when a cluster of human cases in North America was linked to handling fresh fish, particularly tilapia. Clinical signs presented in the human cases were cellulitis and endocarditis, marking the first recognition of *S. iniae* as a significant zoonotic threat [10]. By the early 2000s, additional human infections were reported, reinforcing the need for improved biosecurity measures and increase food safety awareness. During the same period, frequent outbreaks of *S. iniae* and *S. agalactiae* were evident, prompting extensive research into the epidemiology and pathogenesis of streptococcosis. The introduction of vaccines against *Streptococcus* species provided a means of outbreak control, although early vaccines required multiple doses and had variable efficacy depending on environmental and host factors [11].

In 2003, the adoption of PCR-based molecular diagnostics revolutionized pathogen detection and disease management, enabling rapid and accurate identification of *Streptococcus* species [12]. During the 2010s, genomic studies deepened the understanding of *Streptococcus* virulence factors, paving the way for recombinant protein and DNA vaccines that provided more consistent and long-lasting protection [13]. In 2013, CRISPR-Cas9 genetic engineering demonstrated the potential for developing disease-resistant fish strains, representing a significant leap in biotechnology for aquaculture disease control and prevention [14]. Technological advancements such as big data analytics and IoT integration enabled real-time disease monitoring and predictive modelling, further improving disease control and prevention in aquaculture [15].

A critical event occurred in 2015 in Singapore, where an outbreak of *S. agalactiae* sequence type 283 (ST283) in human was linked to the consumption of raw freshwater fish, particularly tilapia. Unlike typical GBS infections, this outbreak affected healthy human adults, causing severe conditions such as septic arthritis, meningitis, and limb amputations [16-18]. The ST283 strain spread across Southeast Asia, prompting public health campaigns emphasizing the importance of consuming thoroughly cooking fish [19,20].

The role of climate change in disease dynamics also became apparent. By 2015, research linked rising water temperatures with increased *Streptococcus* outbreaks, highlighting how environmental factors influence pathogen emergence [21-23]. In 2019, phytotherapy gained wider recognition in commercial aquaculture, promoting sustainable management practices through plant-based treatments as alternatives to antibiotics [24]. However, research into natural antimicrobials had already been conducted in earlier years [25-27] By 2023, *S. agalactiae* ST283 was reported for the first time in Malaysia, once again linked to raw fish consumption, reinforcing the need for continuous vigilance and public health efforts in the affected regions [28].

This historical perspective highlights the evolving strategies to manage *Streptococcus* infections in aquaculture, emphasizing the interplay between scientific advancements, aquaculture practices, and environmental changes.

**Table 1.** Key historical milestones in the study and management of *Streptococcus* infections in aquaculture, highlighting significant events, discoveries, and their impact on industry practices. The timeline includes the first isolation of *Streptococcus* species, notable outbreaks, advancements in diagnostics and vaccines, and the introduction of new management strategies in response to emerging challenges, supported by references to pivotal studies.[23]

| Year | Event/Discovery | Description | Impact on Aquaculture | References |
|---|---|---|---|---|
| 1957 | First report of streptococcal disease in fish | Streptococcal infection documented in cultured rainbow trout (*Oncorhynchus mykiss*) in Japan | First recognition of *Streptococcus* as an aquatic pathogen, initiating research into disease management in aquaculture | [4,5] |
| 1972 | First isolation of *Streptococcus iniae* | Isolated from an Amazon river dolphin (*Inia geoffrensis*), later identified as a major aquaculture pathogen | Marked initial recognition of its zoonotic potential and impact on fish farming | [6] |
| 1980s | Emergence of *Streptococcus agalactiae* (GBS) in aquaculture | Outbreaks reported in tilapia farms across Southeast Asia | Highlighted the expansion of streptococcal infections in warm-water fish farming, compounded by antibiotic resistance | [5] |
| 1990s | Identification of *Streptococcus ictaluri* and *Streptococcus parauberis* | *S. ictaluri* identified as a pathogen in North American catfish, *S. parauberis* affecting turbot and sea bass in Europe | Expanded understanding of *Streptococcus* species in different aquaculture systems, influencing management practices | [7,8] |
| 1995-1996 | First zoonotic outbreak of *Streptococcus iniae* | Cluster of human cases in North America linked to handling fresh fish, particularly tilapia | Raised global awareness of zoonotic risks, leading to increased biosecurity measures and handling guidelines in aquaculture | [10] |
| 1999 | First isolation of *Streptococcus phocae* | Isolated from Atlantic salmon in Chile | Expanded knowledge of *Streptococcus* infections in marine environments | [9] |
| Early 2000s | Development of *Streptococcus* vaccines | Introduction of formalin-killed vaccines for *S. iniae* and *S. agalactiae* | Provided a means of controlling outbreaks, although efficacy varied by species and environmental conditions | [11] |
| 2003 | Adoption of molecular diagnostics | PCR-based techniques introduced for rapid detection of *Streptococcus* in aquaculture | Enabled early detection and more effective management of disease outbreaks | [12] |
| 2005 | *Streptococcus suis* zoonotic outbreak in China | Over 200 human cases, including fatalities, linked to handling or consuming undercooked pork in Sichuan Province | Highlighted the zoonotic risks of *Streptococcus* species beyond aquaculture, reinforcing the need for stringent biosecurity | [29] |
| 2010s | Advances in vaccine development | Genomic studies led to recombinant protein and DNA vaccine development | Provided more consistent and long-lasting protection against *Streptococcus* infections in aquaculture | [30,31] |
| 2013 | Application of CRISPR-Cas9 in aquaculture | Genetic engineering demonstrated potential for creating disease-resistant fish strains | Marked a breakthrough in biotechnology for combating *Streptococcus* infections, offering a new avenue for disease control | [14] |

| 2015 | *Streptococcus agalactiae* ST283 outbreak in Singapore | Major outbreak linked to consumption of raw freshwater fish, affecting healthy adults with severe infections | Prompted significant public health responses and stricter food safety regulations regarding raw fish consumption | 16-18 |
|------|------|------|------|------|
| 2015 | Recognition of climate change impact on *Streptococcus* infections | Studies linked rising water temperatures to increased outbreaks | Reinforced the role of environmental factors in disease dynamics, leading to climate-focused research in aquaculture disease management | 21-23 |
| 2019 | Increased recognition of phytotherapy in aquaculture | Expansion of plant-based treatments as sustainable alternatives to antibiotics in disease management | Encouraged research into natural antimicrobials and reduced reliance on antibiotics in aquaculture | 24-26 |
| 2023 | First report of *Streptococcus agalactiae* ST283 in Malaysia | Linked to raw fish consumption, continuing the spread of ST283 in Southeast Asia | Reinforced the need for ongoing vigilance, surveillance, and public health efforts in affected regions | 28 |

## 2.2 Description of *Streptococcus* Species Affecting Aquaculture

*Streptococcus* infections represent a significant threat to the aquaculture industry, with *S. iniae* and *S. agalactiae* being the predominant species involved. *S. iniae* is a Gram-positive, beta-hemolytic bacterium that was initially isolated from the Amazon freshwater dolphin, *Inia geoffrensis* [32]. It exhibits catalase-negative and oxidase-negative characteristics and grows optimally at temperatures ranging from 25°C to 37°C. *S. iniae* forms chains of spherical cells, and its cell wall structure is characteristic of the Lancefield Group C streptococci [33]. Genetic analysis has revealed significant heterogeneity among *S. iniae* strains. Multi-locus sequence typing (MLST) has identified multiple sequence types (STs), indicating a high level of genetic diversity. Comparative genomic studies have shown variations in virulence genes, including those encoding capsular polysaccharides, surface proteins, and toxins. This genetic diversity complicates the development of effective vaccines and necessitates continuous monitoring [33]. Furthermore, *S. iniae* exhibits resistance to multiple antibiotics, including oxolinic acid, sulphamethoxazole-trimethoprim, and amoxicillin. The resistance mechanisms are often associated with mutations in the quinolone resistance-determining regions (QRDR) of the gyrA and parC genes and the acquisition of resistance genes through horizontal gene transfer. Studies have reported heritability of resistance traits, suggesting that selective breeding could enhance resistance in aquaculture species [3].

*Streptococcus agalactiae*, or Group B Streptococcus (GBS), is a Gram-positive, beta-hemolytic bacterium. It is facultatively anaerobic, catalase-negative, and typically forms chains of cocci. *S. agalactiae* can grow at a wide range of temperatures but prefers 35-37°C. It is classified into several serotypes based on the capsular polysaccharide antigens, with serotypes Ia, Ib, and III being the most relevant in aquaculture [34]. *S. agalactiae* exhibits considerable genetic variability, with different serotypes and sequence types adapted to specific hosts. Genome sequencing has revealed substantial differences in gene content among strains, particularly in the regions encoding surface proteins, virulence factors, and antibiotic resistance genes. Serotype Ib strains, highly adapted to fish, show significant genome reduction compared to other serotypes, reflecting niche specialization [35]. *S. agalactiae* is known for its resistance to multiple antibiotics, including penicillin, erythromycin, tetracycline, and vancomycin. Resistance mechanisms involve alterations in penicillin-binding proteins (PBPs), efflux pumps, and modifications of ribosomal targets. The global trade of aquaculture species, particularly tilapia, has facilitated the spread of resistant strains, necessitating robust surveillance program and the development of novel antimicrobial strategies [36].

## 2.3    Epidemiology of Streptococcosis

### 2.3.1    Geographic distribution

*Streptococcus* spp in aquaculture are distributed globally, affecting various regions and aquatic environments (Figure 1, Table 2). In Southeast Asia, *S. agalactiae*, particularly sequence type (ST) 283, has been reported in significant outbreaks in Thailand, Vietnam, Singapore, and Malaysia [19,37-42]. In Australia, *S. agalactiae* serotype Ib has been found in wild fish and captive stingrays, likely introduced through the importation of tilapia from Israel during the 1970s and 1980s [35]. Outbreaks of *S. agalactiae* has also been reported from fish farms in Iran [43]. In North Africa, *S. agalactiae* serotype IV has been identified in farmed tilapia, marking the first report of this serotype in fish [34]. Similar strains are found in North America [20]. In South America, *S. agalactiae* has also been reported from tilapia farms in Brazil [44]. In the East Asia region, *S. agalactiae* serotype Ia affects tilapia farms in Taiwan, where climatic factors have been shown to influence susceptibility [45]. Additionally, China has reported significant outbreaks of *S. agalactiae* in fish farms [46,47].

*S. iniae* has also been reported in farmed and wild fish species across various regions. In North America, it affects farmed and wild fish in Mexico and the USA [10,48]. The pathogen is also prevalent in the Caribbean affecting both farmed and wild fish [33]. In the Mediterranean Sea region, *S. iniae* has been detected in wild marine fish in Israel [49]. Similarly, the pathogen has been detected in rainbow trout farms in west Iran [50]. In Southeast Asia, *S. iniae* is has been reported in tilapia farms in Malaysia [51], the Philippines [3], Indonesia [38], and Thailand [39,52]. Europe also reports the presence of *S. iniae* in fish farms in Spain and Italy [48,53,54].

The global spread of these infections is significantly influenced by the movement of aquaculture species through international trade routes, environmental conditions, and biosecurity practices [35].

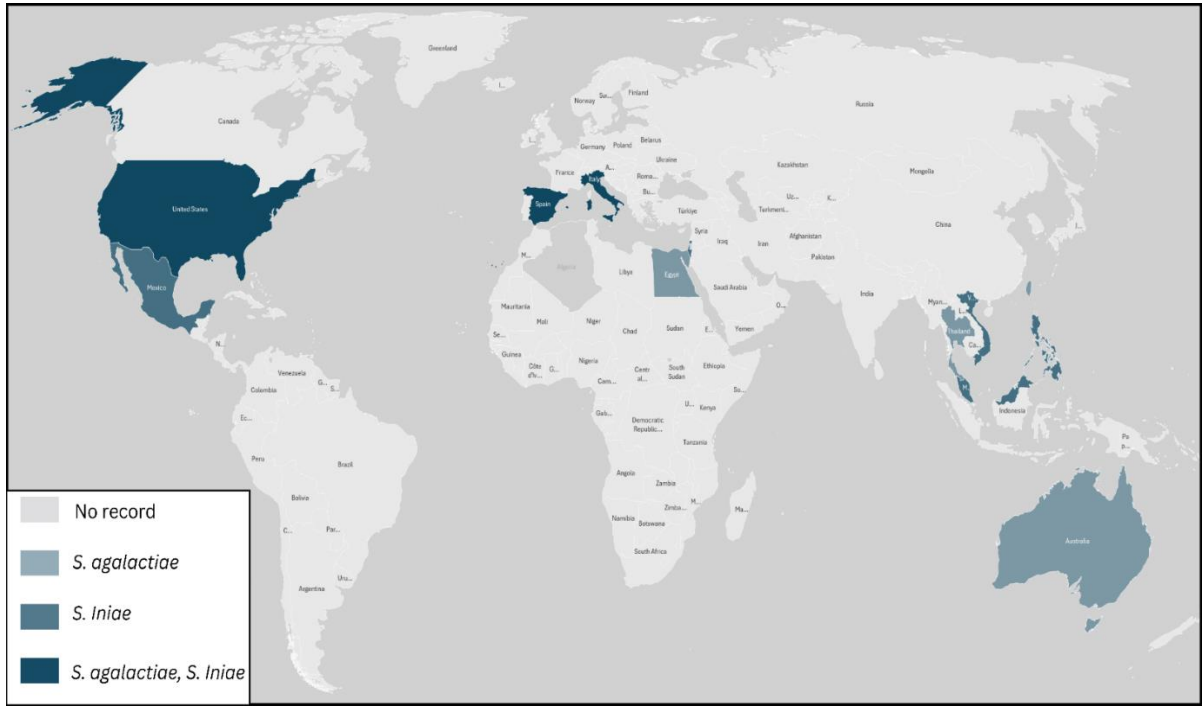

**Figure 1.** Global distribution of *Streptococcus* infections in aquaculture. Different colors represent the presence of *Streptococcus agalactiae* and *Streptococcus iniae* in various countries. Regions where these pathogens have been reported are highlighted, demonstrating the geographical spread and prevalence of *Streptococcus* species in aquaculture across different continents.

**Table 2.** Ccomparative epidemiology of major *Streptococcus* species impacting aquaculture, including their primary hosts, geographic distribution, transmission routes, and clinical manifestations. The table also highlights mortality rates associated with outbreaks, emphasizing the severity of infections under varying conditions. References to key studies are included to support the documented information

| *Streptococcus* species | Affected Host Species | Geographic Distribution | Transmission Routes | Primary Clinical Manifestation | Associated Mortality Rates | References |
|---|---|---|---|---|---|---|
| *Streptococcu. agalactiae* | Tilapia, Catfish, Rainbow Trout, Crucian Carp, Ya-fish, Golden Pomfret, Frogs, Wild Fish, Stingrays | Southeast Asia (Thailand, Vietnam, Singapore, Malaysia), Australia, North Africa, North America, South America, China, Taiwan | Oral (via ingestion of contaminated water or feed), Handling of infected fish | Lethargy, exophthalmia, hemorrhages, neurological signs, high mortality rates | Can exceed 50% in severe outbreaks | [5,16,17,19,21,34,36] |
| *S. iniae* | Tilapia, Channel Catfish, Rainbow Trout, Sea Bass, Sea Bream, Asian Sea Bass, Red Drum, Wild Fish, Crustaceans | North America (USA, Mexico), Caribbean, Mediterranean (Israel), Southeast Asia (Malaysia, Philippines, Indonesia, Thailand), Europe (Spain, Italy) | Direct contact, Handling of infected fish, Waterborne transmission | Lethargy, erratic swimming, hemorrhages, meningoencephalitis, high mortality | Varies, often 20-50% in outbreaks | [3,6,10,33,48,51] |
| *S. ictaluri* | Catfish | Southern USA | Waterborne, Direct contact | High mortality, systemic infection, organ damage | Can reach up to 40% during outbreaks | [7] |
| *S. parauberis* | Turbot, Sea Bass | Europe | Waterborne, Vertical transmission | Granulomas, abscesses, systemic infection | Moderate to high, depending on environmental factors | [8] |
| *S. phocae* | Atlantic Salmon | Chile | Direct contact, Waterborne transmission | Ulcerative lesions, systemic infection | Often high in intensive aquaculture systems | [9] |
| *S. suis* | Pigs (relevant due to zoonotic potential) | China | Direct contact with infected animals, Ingestion of contaminated food | Meningitis, Septicemia, Arthritis | High mortality in untreated human cases | [29] |

### 2.3.2 Host species affected

*Streptococcus* infections in aquaculture affect a diverse range of host species across freshwater and marine environments (Table 2). *S. agalactiae* primarily impacts farmed fish, with tilapia (*Oreochromis spp.*) being highly susceptible, particularly serotypes Ia, Ib, III, and ST283, which have caused severe outbreaks [19,42]. Hybrid tilapia are notably affected by serotypes Ia ST7 and III ST283 [42]. Catfish (*Ictalurus spp.*), though less frequently infected, experience high morbidity and mortality during outbreaks [36]. Rainbow trout (*Oncorhynchus mykiss*) face significant losses due to *S. agalactiae*, leading to reduced fish quality and survival rates [43]. Other hosts include crucian carp (*Carassius carassius*), ya-fish (*Schizothorax prenanti*), golden pomfret (*Trachinotus blochii*), frogs, wild fish, and stingrays, demonstrating the pathogen's adaptability [35,41,47,55,56].

*S. iniae* also infects a broad spectrum of species, with tilapia being the primary host, suffering substantial economic losses from high mortality rates [3]. Hybrid tilapia experience severe disease outbreaks due to *S. iniae* [51]. Channel catfish (*Ictalurus punctatus*), rainbow trout, and marine species like sea bass (*Dicentrarchus labrax*) and sea bream (*Sparus aurata*) are significantly affected [33,49,50]. Asian sea bass (*Lates calcarifer*) has shown increased susceptibility, while red drum (*Sciaenops ocellatus*), wild fish, and crustaceans further illustrate the pathogen's wide host range [10,52]. The extensive host range of *Streptococcus* spp. emphasizes need for precise disease control and management strategies to mitigate its impact in aquaculture.

### 2.3.3 Incidence and prevalence rates

*Streptococcus* spp in aquaculture present significant risks towards aquatic animal with varying incidence and prevalence rates influenced by region, species, and environmental conditions. In the Levantine Basin of the Mediterranean Sea, *Streptococcus* spp. were found in 9.71% of wild marine fish and crustaceans, with *S. iniae* detected at a higher prevalence in kidney tissue compared to liver tissue [49]. In Egypt, the prevalence of bacterial infections in farmed Nile tilapia (*Oreochromis niloticus*) was 26.2%, with *S. agalactiae* being the most prevalent at 15.5%, particularly during the summer season [57]. Another Egyptian study reported emerging pathogens, including *S. agalactiae* and *Streptococcus faecalis*, with significant antibiotic resistance observed [58]. Additionally, *S. agalactiae* caused a 34.9% prevalence in red tilapia mortalities during the summer [59].

In the United States, *S. iniae* was found in 3.81% of tilapia and 7.23% of hybrid striped bass (*Morone chrysops × Morone saxatilis*) on commercial fish farms, with the highest prevalence during the grow-out stage [60]. In China, a shift was observed between 2006 and 2011, where *S. iniae* was gradually replaced by *S. agalactiae* as the dominant tilapia pathogen [46]. In Malaysia, *S. agalactiae* prevalence was significantly higher in lake environments compared to rivers, correlating strongly with increased fish mortalities [21].

In Southeast Asia, particularly Thailand and Indonesia, *Streptococcus* infections are widespread in farmed fish, especially tilapia and Asian sea bass. In Thailand, a high prevalence of *S. agalactiae* and *S. iniae* infections was reported, with 86.67% of samples testing positive for *S. agalactiae*, 8.48% for *S. iniae*, and 4.85% as mixed infections [39]. *S. iniae* has been isolated from Asian sea bass (*Lates calcarifer*) and red tilapia, with higher virulence observed in Asian sea bass [52]. In Indonesia, infections with *S. iniae*, *S. agalactiae*, and *Lactococcus garvieae* were identified in Nile tilapia cultured in net cages [38]. Outbreaks of *S. agalactiae* and *S. iniae* in both Thailand and Indonesia resulted in 40-60% mortality rates in floating net cages and continuous lower daily mortalities in pond systems [61].

In Mexico, *S. iniae* caused significant outbreaks in two geographically isolated tilapia populations, with mortality rates reaching 68% and 80% in different outbreaks [48]. The economic burden of *Streptococcus* infections is significant, as evidenced by the USD $250 million loss in tilapia production in 2006 [48]. These varying prevalence rates, as presented in Table 2, highlight the widespread and persistent nature of *Streptococcus* infections in global aquaculture, emphasizing the need for effective disease management strategies.

### 2.3.4   Zoonotic Risks Associated with Streptococcus Infections

*Streptococcus* spp, particularly *Streptococcus iniae* and *Streptococcus agalactiae*, poses a significant public health concern, especially in regions with intensive aquaculture. These pathogens can infect and lead to severe illnesses in humans. The risk of zoonotic transmission is heightened by close human interaction with aquatic environments and cultural practices involving the handling and consumption of raw or undercooked fish (ref?).

One of the earliest recognized human outbreak of *S. iniae* occurred in North America during the winter of 1995-1996. In the Greater Toronto area, several individuals developed invasive clinical signs such as cellulitis and endocarditis after handling fresh fish, particularly tilapia.

These infections were traced to skin injuries sustained during fish handling, highlighting the risks posed by direct contact with infected fish [10]. Similar cases were reported in Hong Kong in 2003, where patients who regularly handled fresh fish developed septic arthritis and bacteremic cellulitis, with regional variations in *S. iniae* virulence observed between North America and Asia [62].

In Southeast Asia, *S. agalactiae* sequence type 283 (ST283) has emerged as a significant zoonotic pathogen. The first major outbreak occurred in Singapore in 2015, affecting 146 individuals who consumed raw freshwater fish, predominantly tilapia. This outbreak was unusual in that it primarily affected healthy adults, leading to severe infections such as septic arthritis and meningitis, with some cases requiring limb amputations [20,63]. The hypervirulent ST283 strain has since spread across Southeast Asia, including Malaysia and Laos, emphasizing the regional impact and public health challenges posed by the cross-border movement of fish and people. In Malaysia, the first human cases were reported in 2023, linked to the consumption of raw freshwater fish [28]. Similarly, in Lao PDR, a report documented the simultaneous occurrence of invasive *S. agalactiae* ST283 infection in two sisters who had consumed raw freshwater fish, resulting in sepsis in otherwise healthy adults [17].

Beyond aquaculture, *Streptococcus suis*, primarily a pathogen of pigs, has caused significant zoonotic infections in humans, particularly in China. A notable outbreak in Sichuan province in 2005 led to over 200 cases of severe infections, including meningitis and septicemia, with more than 30 fatalities. This outbreak, linked to handling or consuming undercooked pork, highlights the broader zoonotic risks associated with *Streptococcus* species in regions with intensive animal farming and aquaculture [29]. Although *S. suis* is primarily associated with pigs, its zoonotic relevance in aquaculture is recognized due to similar transmission routes and shared public health risks.

Table 3 provides an overview of documented zoonotic *Streptococcus* outbreaks, outlining transmission routes, clinical manifestations, and public health responses. These outbreaks highlight the critical role of food safety regulations, public health education, and biosecurity measures in mitigating zoonotic risks.

Aquaculture workers and fish processors face heightened risks, especially in regions where raw fish consumption is prevalent. The severity of these infections, particularly among vulnerable groups such as the elderly and immunocompromised individuals, reinforces the need for proactive disease control and prevention. The *Streptococcus agalactiae* ST283 outbreak in Singapore exemplifies how contaminated seafood can pose significant public health threats, emphasizing the importance of continuous surveillance and rapid response measures. To minimize these public health threats, a comprehensive one health approach is necessary. Raising public awareness about safe seafood handling and preparation, enforcing stringent food safety regulations and strengthening biosecurity measures on fish farms are essential. In regions with high consumption of raw fish, targeted disease control and prevention can help prevent future outbreaks, and the interventions not only could improve the sustainability of the aquaculture industry but also ensuring public health protection.

**Table 3.** Documented zoonotic outbreaks of *Streptococcus* species across various regions and years, detailing the transmission routes, clinical manifestations in humans, and public health responses. Public health impacts are highlighted, emphasizing the importance of food safety regulations, public health education, and biosecurity measures in preventing and managing zoonotic risks associated with *Streptococcus* infections. References to key studies provide further context for each incident.

| Year | Region/Country | Streptococcus Species | Transmission Route | Clinical Manifestations in Humans | Public Health Response | Public Health Impact | References |
|---|---|---|---|---|---|---|---|
| 1995-1996 | Canada (Greater Toronto Area) | *Streptococcus iniae* | Handling of infected fish (*tilapia*) | Cellulitis, Endocarditis, Meningitis | Public health alert, enhanced surveillance of fish markets, recommendations for safe fish handling practices. | Raised awareness of zoonotic risks associated with fish handling; influenced handling practices in aquaculture | [10] |
| 2003 | Hong Kong | *Streptococcus iniae* | Handling of infected fish | Septic arthritis, Bacteremic cellulitis | Public health advisory, monitoring of fish markets, implementation of educational programs for fish handlers | Enhanced awareness of zoonotic risks in fish markets; influenced safety protocols in fish handling and processing | [62] |
| 2005 | China (Sichuan Province) | *Streptococcus suis* | Handling and consumption of undercooked pork | Septicemia, Meningitis, Arthritis | Quarantine measures, public health campaigns to avoid undercooked pork, development of vaccination programs for pigs | Over 200 human cases, more than 50 deaths; led to stricter food safety regulations and improved animal husbandry practices | [29] |
| 2015 | Singapore | *Streptococcus agalactiae* (ST283) | Consumption of raw freshwater fish (*tilapia*) | Septicemia, Meningitis, Arthritis | Ban on sale of raw freshwater fish, extensive public health campaigns, development of guidelines for safe fish consumption | 146 human cases; led to significant changes in food safety regulations and increased public awareness of risks associated with raw fish consumption | [16,63] |
| 2015 | Thailand | *Streptococcus agalactiae* (ST283) | Consumption of raw freshwater fish | Septicemia, Meningitis | Public health advisory, reinforcement of food safety regulations, enhanced monitoring of fish products | Increased international attention to zoonotic risks; reinforced need for safe fish consumption practices | [19] |
| 2019 | Lao DPR | *Streptococcus agalactiae* (ST283) | Consumption of raw freshwater fish | Septicemia, Meningitis | Strengthened food safety regulations, public health education, introduction of vaccination strategies in pigs | Decreased incidence of zoonotic transmission; highlighted the effectiveness of combined public health and veterinary interventions | [17] |
| 2023 | Malaysia | *Streptococcus agalactiae* (ST283) | Consumption of raw freshwater fish | Septicemia, Meningitis | Public health campaigns, food safety advisories, ongoing surveillance | Continued public health concern, highlighted need for | [28] |

| | | | | | regional cooperation in food safety | |
|---|---|---|---|---|---|---|

# 3.0    Pathogenesis and clinical manifestations

## 3.1    Pathogenesis of Streptococcus infections in fish

### 3.1.1    Infection mechanisms

The spread of *Streptococcus* infections in aquaculture occurs through multiple pathways, including direct transmission between fish, vertical transmission from broodstock, and contamination via water and equipment. These transmission routes are visually represented in Figure 2, which illustrates how *Streptococcus* spreads within aquaculture environments and the key factors influencing its persistence and dissemination. Infected fish shed bacteria into the water, contributing to horizontal transmission within the farming system. Additionally, contaminated equipment, handling practices, and water sources act as vectors, exacerbating the risk of infection spread.

*Streptococcus* spp can enter the fish through entry via gills, skin abrasions, and the gastrointestinal tract [64]. The stepwise progression of *S. iniae* infection, from adhesion and invasion to immune evasion and systemic dissemination, is visually represented in Figure 2. This schematic diagram highlights key pathogenesis mechanisms, including entry through epithelial barriers, intracellular survival, immune evasion, and eventual central nervous system infection.

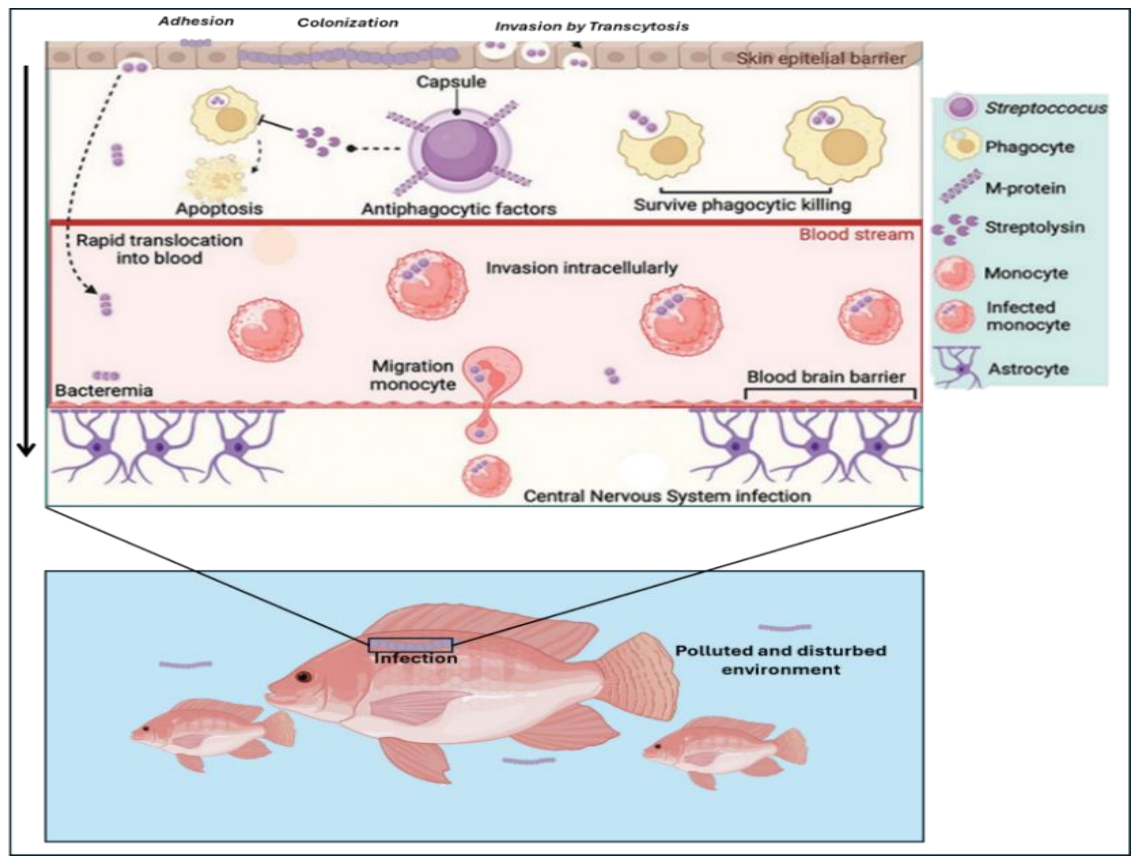

**Figure 2:** Schematic representation of the complex process of *Streptococcus* infection in fish, with a focus on how environmental stressors contribute to the disease's progression. The pathogenesis unfolds in four primary stages: (1) Adhesion – where the pathogen attaches to the fish's epithelial cells; (2) Invasion via Transcytosis – involving the penetration of the epithelial barrier; (3) Immune Evasion and Systemic Spread – the bacteria bypass the immune defenses and enter the bloodstream; and (4) Proliferation and Dissemination – the bacteria multiply and spread within the host, ultimately leading to severe systemic infection and central nervous system involvement. Figure adapted from Juarez-Cortes et al., (2024).

*Streptococcus* spp could adhere to epithelial cells using surface proteins and adhesins, which facilitate colonization and proliferation by evading the host's immune responses [65-67]. Initially, *Streptococcus* spp. produce extracellular enzymes and toxins that degrade host tissues, aiding bacterial invasion [68]. For instance, *S. agalactiae* produces hemolysins that lyse red blood cells, releasing nutrients to support bacterial growth [69,70]. Enzymes like hyaluronidase and proteases further break down connective tissues, facilitating dissemination within the host [46]. To evade host immune defenses, *Streptococcus* spp. form capsules, preventing phagocytosis by immune cells [71]. The polysaccharide capsule masks bacterial antigens, making it difficult for the immune system to recognize and attack the pathogens [72,73]. Additionally, *Streptococcus* spp. secrete proteins that inhibit the complement system, a crucial component of the innate immune response, enhancing bacterial survival [74]. Once established, the bacteria spread to various tissues and organs, leading to systemic infection. The bloodstream acts as a conduit for bacterial dissemination, causing widespread damage [74]. Infected fish often exhibit septicaemia, with bacteria present in the blood and severe systemic inflammation. This systemic spread results in multiple organ dysfunctions, contributing to high mortality rates in aquaculture populations [50,58].

The role of environmental stressors in disease progression is also emphasized, demonstrating how pollution and compromised host immunity exacerbate infections (Figure 3). *Streptococcus* spp. can form biofilms on mucosal surfaces and other tissues, providing a protective environment that shields them from the host's immune system and antibiotic treatments [75]. Biofilm formation is particularly problematic in aquaculture settings, facilitating persistent infections and complicating eradication efforts [76,77]. *S. iniae* can invade and survive within host cells, efficiently targeting macrophage-like cells and pronephros phagocytes, allowing persistence and multiplication [12,78,79]. *S. iniae* type II survives within phagocytes for at least 48 hours and induces apoptosis, enhancing its ability to cause systemic infection [80]. Additionally, *S. iniae* produces streptolysin S (SLS), a pore-forming cytotoxin crucial for its virulence, with mutants lacking SLS showing significantly reduced virulence [73,80]. These mechanisms highlight the complexity of *Streptococcus* infections in fish, emphasizing the pathogen's ability to evade the host's immune system and cause widespread systemic damage. A comparative overview of infection mechanisms across *Streptococcus* species, including their entry points, virulence factors, and clinical outcomes, is provided in Table 4.

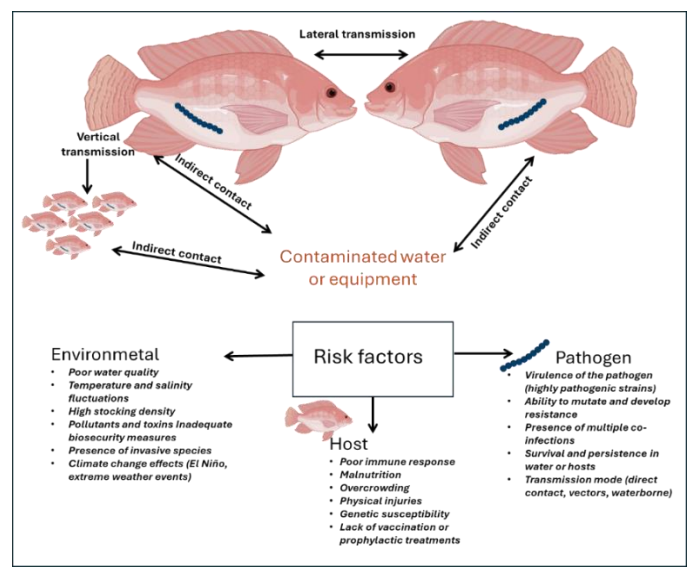

**Figure 3:** Transmission pathways of *Streptococcus* infections in aquaculture. The bacteria can be transmitted through direct fish-to-fish contact, via skin abrasions and gill exposure. Contaminated water and equipment serve as major vectors, as bacteria cells persist in water bodies, biofilms, and farm tools, facilitaing horizontal transmissison within aquaculture systems. Infected fish can also shed *Streptococcus* spp. into the surrounding water, furhter increasing pathogen load and exposure for healthy individuals.

**Table 4.** Pathogenesis of various *Streptococcus* species affecting aquaculture, including their entry points, disease progression, key virulence factors, clinical manifestations, and clinical outcomes. This table outlines the mechanisms of infection, including bacterial entry routes, virulence strategies, and resulting symptoms in affected fish. It also highlights the severity of infections and references key studies documenting these pathogenic processes.

| Streptococcus Species | Entry Points | Disease Progression | Key Virulence Factors | Symptoms | Clinical Outcomes | References |
|---|---|---|---|---|---|---|
| *Streptococcus iniae* | Skin abrasions, Gills | Initial colonization of the skin/gills, rapid systemic spread via bloodstream | Capsular polysaccharides: Prevents phagocytosis; Hemolysins: Causes lysis of red blood cells; M-proteins: Interferes with immune response | Disorientation, erratic swimming, skin lesions, lethargy | Septicemia, Meningitis, Ulcerative lesions, High mortality in severe cases | 12,73,78-80 |
| *Streptococcus agalactiae* (ST283) | Oral ingestion, Gills | Colonization of the gastrointestinal tract and gills, followed by systemic dissemination | Hemolysins: Causes tissue damage; Hyaluronidase: Facilitates tissue invasion; Fibrinogen-binding proteins: Aids in immune evasion | Lethargy, anorexia, abnormal swimming, joint swelling (arthritis) | Septic arthritis, Meningitis, High mortality | 16,17,46,69,70 |
| *Streptococcus parauberis* | Gills, Mucous membranes | Localized infection at the gills, leading to systemic invasion | Exopolysaccharides: Protects against immune response; Surface proteins: Involved in adherence and colonization | Respiratory distress, lethargy, abdominal swelling (peritonitis) | Peritonitis, Septicemia, Meningitis, High mortality if untreated | 8,74 |
| *Streptococcus phocae* | Gills, Skin | Colonization of skin/gills, rapid systemic spread via bloodstream | Hemolysins: Induces cell lysis; Capsular polysaccharides: Prevents immune clearance | Neurological signs, lethargy, abnormal swimming, skin lesions | Meningitis, Septicemia, High mortality in farmed salmon | 9 |
| *Streptococcus ictaluri* | Skin lesions, Gills | Initial infection at skin lesions or gills, with bacteria spreading to internal organs | Proteases: Degrade host tissues; Lipoteichoic acid: Facilitates adherence to host cells | Ulceration, internal hemorrhage, lethargy, loss of appetite | Ulcerative disease, Septicemia, Significant mortality in catfish | 7,50 |
| *Streptococcus suis* | Oral ingestion, Skin wounds (Zoonotic) | Entry through oral ingestion or skin wounds, with rapid systemic spread, meningitis development | Capsule: Provides resistance to phagocytosis; Muramidase-released protein (MRP): Enhances bacterial survival; Extracellular factor (EF): Contributes to virulence | Neurological signs (incoordination, convulsions), joint swelling, fever | Meningitis, Septicemia, Arthritis in pigs; Zoonotic transmission to humans | 29 |

### 3.1.2 Host-pathogen interactions

Host-pathogen interactions in *Streptococcus* infections involve a dynamic interplay of bacterial adhesion, immune evasion mechanisms, intracellular survival, apoptosis induction, virulence factors, and host immune responses. Upon infection, the innate immune system is the first line of defense, with mucosal barriers, phagocytic cells (macrophages, neutrophils), and reactive oxygen species (ROS) production playing crucial roles [81].

Following entry, *Streptococcus* spp. adhere to host cells using various surface proteins and adhesins. For instance, *S. iniae* employs the simA gene to encode an M-like protein, which enhances adhesion, invasion, and resistance to phagocytic killing [82]. Similarly, *S. dysgalactiae* displays high surface hydrophobicity and hemagglutinating activity, facilitating strong adherence to carp epithelial cells [83].

To evade immune responses, *Streptococcus* spp. produce polysaccharide capsules, which inhibit phagocytosis and complement system activation, thus enhancing bacterial survival [84]. The capsule operon in *S. iniae* regulates capsule synthesis and length, further strengthening its immune evasion capabilities [84]. In addition, proteins secreted by *Streptococcus* spp., such as C5a peptidase and interleukin-8 protease, actively disrupt host immune signaling pathways, hindering the recruitment and activation of phagocytes [73,85].

The virulence mechanisms of *S. iniae*, including capsule formation, streptolysin production, chemokine degradation, and fibrin clot breakdown, are schematically represented in Figure 4. This figure highlights the mgx gene-driven regulation of SiM protein, which enables bacterial adhesion to immunoglobulins and fibrinogen, aiding in host immune evasion. It also depicts the role of streptolysin S, controlled by the sivS/R regulatory system, in targeting lymphocytes, neutrophils, and erythrocytes, further compromising host immune defenses.

Certain strains of *Streptococcus* spp. also exhibit intracellular survival strategies. *S. iniae* can invade macrophage-like cells and pronephros phagocytes, persisting and multiplying within the host. *S. iniae* type II survives inside phagocytes for at least 48 hours, inducing apoptosis of immune cells, which weakens the host response and facilitates systemic infection [12].

Additionally, *Streptococcus* spp. utilize various virulence factors that contribute to their pathogenicity. *S. iniae* produces streptolysin S (SLS), a pore-forming cytotoxin crucial for its virulence, with mutants lacking SLS exhibiting significantly reduced virulence [86]. Similarly, *S. agalactiae* produces hemolysins that lyse red blood cells, releasing essential nutrients for bacterial proliferation [50]. Other enzymes, such as hyaluronidase and proteases, break down connective tissues, facilitating bacterial dissemination within the host [46]. These virulence factors and their contributions to disease progression are outlined in Table 4.

The host immune response to *Streptococcus* infections involves both innate and adaptive immunity. The innate response includes phagocytosis, ROS production, and cytokine secretion (e.g., IL-1β, IL-6, IL-8, TNF-α), all of which play critical roles in mounting an effective immune defense [87,88]. Meanwhile, adaptive immunity involves antibody production and T-cell activation, which are essential for long-term immunity and pathogen clearance. Additionally, nutritional immunity, wherein the host limits iron availability, serves as an important mechanism to control bacterial growth. However, *Streptococcus* spp. counteract this through iron-acquisition mechanisms, such as siderophores and heme-utilization systems [83].

These intricate interactions between the host and the pathogen lead to various clinical manifestations, including tissue damage, systemic inflammation, and septicemia. For instance, *S. agalactiae* infections in tilapia commonly present as hemorrhages, exophthalmia (pop-eye), and organ enlargement [89].

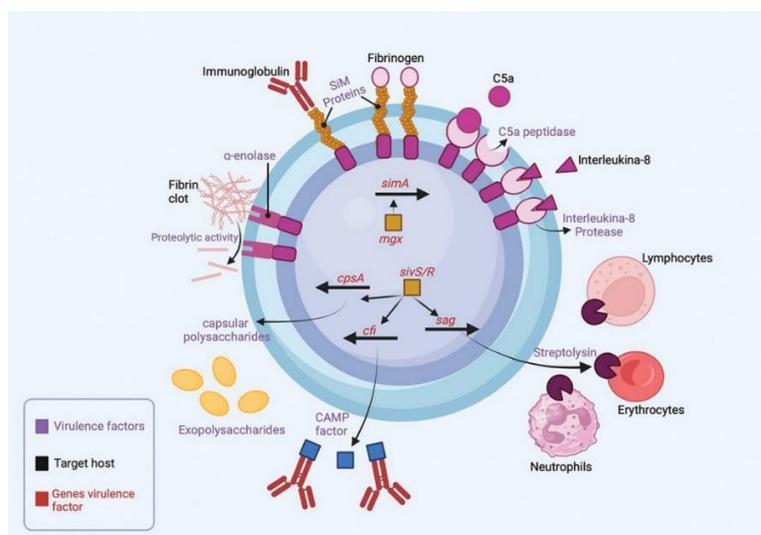

**Figure 4.** Schematic illustration of the complex interplay of virulence factors in *Streptococcus iniae*, emphasizing both intracellular regulatory genes and extracellular factors critical to the bacterium's pathogenicity. Central to this process is the *mgx* gene, which drives the production of the SiM protein (*simA*), enabling the bacterium to bind key host molecules such as immunoglobulin and fibrinogen. The schematic also highlights the role of peptidase C5a and interleukin-8 protease in disrupting host immune signaling by degrading chemokines, thereby impairing the function of phagocytes. Additionally, the *sivS/R* regulatory system is shown to control the production of streptolysin S, a potent cytotoxin that targets lymphocytes, erythrocytes, and neutrophils, as well as the CAMP factor gene (*cfi*), which facilitates immunoglobulin binding. The synthesis of capsular polysaccharides, regulated by *cpsA* under the influence of *sivS/R*, is crucial for immune evasion, while the overproduction of exopolysaccharides contributes to the bacterium's viscous growth. Furthermore, α-enolase is depicted as a key enzyme in breaking down fibrin clots, aiding in the dissemination of the bacteria within the host. Figure originally published in Juarez-Cortes et al., (2024).

## 3.2 Clinical signs in infected fish

### 3.2.1 External and internal symptoms

Streptococcosis in fish manifest through a variety of external and internal symptoms that indicate severe systemic infection. The severity and type of external symptoms can vary depending on the *Streptococcus* species involved, the fish species affected, and the environmental conditions. Infected fish often exhibit abnormal swimming behaviors such as listlessness, erratic movements, and serpentine swimming patterns, which are early indicators of distress and systemic infection [90]. Common external signs include darkened coloration, accompanied by hemorrhages on the skin, fins, and mouth due to vascular damage and bacterial invasion [91]. Exophthalmia, or "pop-eye", characterized by the protrusion of one or both eyes, results from fluid accumulation and inflammation behind the eyes [37]. Additionally, skin ulcers and lesions are frequently observed, which can lead to secondary infections and further complications [92]. Fin and tail rot due to necrosis and bacterial invasion is another significant symptom, often seen in advanced stages of the infection [17]. Other external symptoms include corneal opacity and petechiae around the mouth and fins, indicating severe vascular involvement [93].

Internally, *Streptococcus* infections are marked by ascites, where fluid accumulates in the abdominal cavity, leading to abdominal distension [94]. Infected fish often show signs of organomegaly, particularly hepatomegaly and splenomegaly, due to inflammation and bacterial proliferation in these organs [92]. Internal hemorrhages in the liver, kidney, and heart, along with severe inflammation and necrosis, are common findings [55]. Histopathological examinations frequently reveal multifocal whitish foci in various organs, indicative of abscesses and granulomas [95]. These granulomas are often found in the liver, kidney, spleen, and sometimes the heart, showing extensive tissue damage and immune response to infection. Significant liver damage is also a hallmark of *Streptococcus* infections, characterized by hepatocellular degeneration and necrosis [2]. Additionally, severe inflammatory responses, including necrotizing inflammation in multiple organs such as the liver, kidney, heart, and brain, are commonly observed [96]. Brain histopathology often reveals extensive damage, particularly in areas controlling swimming activities, leading to neurological symptoms such as meningitis and meningoencephalitis [97]. These internal symptoms contribute to high mortality rates, severely impacting aquaculture operations. The clinical manifestations of different *Streptococcus* species, including specific symptoms, progression patterns, and clinical outcomes, are summarized in Table 4.

### 3.2.2 Diagnostic methods

The diagnosis of *Streptococcus* infections in fish has significantly advanced, evolving from traditional culture-based methods to sophisticated molecular techniques. Traditional diagnostic methods such as bacterial isolation, biochemical tests, and serological assays remain fundamental. These methods involve culturing samples on selective media and identifying the bacteria based on colony morphology and biochemical characteristics. Histopathology and immunohistochemistry have also been utilized to observe tissue changes and immune responses, aiding in the diagnosis of streptococcal infections [71,98].

Molecular diagnostic techniques have revolutionized the detection and identification of *Streptococcus* species in fish. Among these, 16S rRNA sequencing is highly regarded for its reliability and specificity, allowing for the precise identification of bacterial species. This method has been a staple in differentiating various *Streptococcus* species and understanding their genetic relationships [71,99]. Multiplex PCR and qPCR assays have been developed to target specific genes, such as the lactate oxidase gene and the 16S-23S rRNA intergenic spacer region, enhancing the detection sensitivity and specificity for pathogens like *S. agalactiae*, *S. iniae*, and *S. dysgalactiae* [100,101]. Nonlethal sampling methods have also been standardized, providing a practical approach for routine health monitoring in aquaculture. Techniques such as kidney aspiration, venipuncture, gill mucus swabs, and fecal sample collection allow for the collection of samples without sacrificing the fish, making them ideal for detecting carrier states and conducting surveillance programs [102]. These samples can be analyzed using bacteriological, PCR, and qPCR methods, ensuring high sensitivity and accuracy in detecting Streptococcus infections.

Fluorescence in situ hybridization (FISH) is another advanced technique used for the rapid identification of *Streptococcus* species. FISH utilizes fluorescent probes that bind to specific ribosomal RNA sequences, allowing for the visualization and identification of bacteria directly in tissue samples. This method is particularly advantageous as it bypasses the need for culture, providing immediate results and enabling rapid intervention [103]. The FTA elute card combined with visual colorimetric loop-mediated isothermal amplification (FTA-e/LAMP) offers a rapid, sensitive, and field-friendly diagnostic tool for detecting *S. agalactiae.* This method involves extracting DNA using the FTA card and then performing a LAMP reaction, which can be visually confirmed, providing a quick and efficient diagnostic solution for aquaculture settings [104].

Moreover, proteomic and molecular fingerprinting techniques, such as MALDI-TOF-MS and REP-PCR, have enhanced the diagnostic capability by providing species-specific identification and strain typing. These methods offer high accuracy and discriminatory power, essential for tracking infection sources and understanding epidemiological patterns [105]. Advancements in molecular, proteomic, and rapid diagnostic methods have significantly improved the diagnosis of Streptococcus infections in fish, enabling early and accurate detection. These tools support ongoing surveillance, health monitoring, and timely interventions, helping to manage and control streptococcosis and mitigate its economic impact on aquaculture.

# 4.0 Current streptococcosis management strategies

## 4.1 Preventive measures

### 4.1.1 Biosecurity practices

Effective biosecurity practices are fundamental for managing and preventing *Streptococcus* infections in aquaculture. At the farm level, biosecurity measures include strict quarantine protocols, egg disinfection, movement control of equipment and personnel, water treatments, and proper disposal of mortalities. Quarantine of new stock and egg disinfection are essential to prevent the introduction of pathogens. Traffic control, including the restriction of personnel and equipment movement, minimizes the risk of cross-contamination between different sections of the farm [21]. Water treatments, such as UV sterilization and ozone treatment, ensure optimal water quality, reducing pathogen loads in aquaculture systems [71]. Proactive measures such as the use of clean feed and proper disposal of mortalities further contribute to reducing pathogen spread and maintaining a healthy aquaculture environment [93]. A comparative overview of biosecurity measures employed in aquaculture, along with their effectiveness, cost-effectiveness considerations, and long-term impact, is summarized in Table 5.

In addition to physical and procedural measures, the use of biosurfactants derived from marine bacteria offers an eco-friendly alternative to antibiotics. These natural compounds disrupt biofilms formed by pathogenic bacteria, reducing their ability to colonize and cause disease [106]. This approach not only strengthens biosecurity but also reduces the over-reliance on antibiotics, thereby mitigating antibiotic resistance risks. Furthermore, WASH (Water, Sanitation, and Hygiene) interventions play a critical role in reducing infection burdens and minimizing antibiotic usage. Maintaining high sanitation standards, including regular cleaning and disinfection of equipment and facilities, significantly prevents pathogen spread [76].

Additional to comprehensive biosecurity programs, incorporation of good husbandry practices, disease surveillance, and improved diagnostic tools can effectively improve the decision making by farmers to prevent, control, and eradicate diseases in aquaculture. Nonlethal sampling methods combined with molecular diagnostic techniques like 16S rRNA sequencing allow early pathogen detection and rapid intervention, further enhancing biosecurity effectiveness [71]. Successful implementation of biosecurity measures depends on education and training of farm personnel to ensure adherence to protocols and to update them on the latest disease management strategies.

**Table 5.** Overview of various disease management strategies employed to control and prevent *Streptococcus* infections in aquaculture. Each strategy is detailed with descriptions of its effectiveness, challenges, cost-effectiveness considerations, long-term impact, and implementation guidelines. Case studies and examples are included to illustrate the practical application of these strategies, supported by references to relevant studies that document their outcomes and implications for aquaculture practices.

| Disease Management Strategy | Description | Effectiveness | Challenges | Cost-benefit Considerations | Long-Term Impact | Implementation Guidelines | Case Studies/Examples | References |
|---|---|---|---|---|---|---|---|---|
| Vaccination | Use of formalin-killed or live-attenuated vaccines to prevent *Streptococcus* infections | High effectiveness in reducing disease incidence, particularly for *S. agalactiae* and *S. iniae* in controlled environments | Variability in efficacy across species and regions; vaccine strain matching is crucial; logistical challenges in administration | High upfront cost but cost-effective in the long term due to reduced mortality and morbidity | Sustainable; reduces reliance on antibiotics and improves overall fish health | Ensure proper vaccine storage and handling; vaccinate during early life stages for best results | Vaccination of tilapia against *S. agalactiae* in Southeast Asia significantly reduced outbreaks | 11,107-110 |
| Antibiotic Treatment | Use of antibiotics like amoxicillin, oxytetracycline, florfenicol, and erythromycin to treat *Streptococcus* infections | Effective in controlling active infections, especially when administered early | Development of antibiotic resistance; environmental concerns; regulatory restrictions; residue issues | Cost varies; short-term solution; long-term costs may increase due to resistance | Unsustainable; growing antibiotic resistance; potential environmental impacts | Use antibiotics as a last resort; monitor for resistance; ensure proper withdrawal times | Use of florfenicol in tilapia farming in China to control *S. agalactiae* outbreaks | 60,111-114 |
| Antimicrobial Peptides (AMPs) | Use of naturally occurring or synthetic peptides with broad-spectrum antimicrobial activity to combat *Streptococcus* infections | Effective against various *Streptococcus* species, including those resistant to antibiotics | Challenges in large-scale production, stability in aquaculture environments, potential toxicity, regulatory hurdles | Initial costs high, but potentially cost-effective in reducing reliance on antibiotics and resistance | Sustainable; could significantly reduce antibiotic use and resistance development | Careful selection of effective AMPs; ensure stability and proper delivery methods | Natural antimicrobial compounds from plant extracts have been tested as alternatives to antibiotics | 115,116 |
| Biosecurity Measures | Implementation of strict hygiene, quarantine, and disinfection protocols to prevent the introduction and spread of pathogens | Highly effective in preventing outbreaks of all *Streptococcus* species when rigorously applied | Requires continuous monitoring; high initial costs; gaps in compliance can lead to failures | High initial investment; cost-effective long-term due to prevention of outbreaks | Sustainable; reduces the need for antibiotics and other treatments; enhances overall farm health | Regular training for staff; establish routine monitoring and compliance checks | Adoption of biosecurity measures in Malaysian tilapia farms reduced *S. iniae* outbreaks | 117 |
| Phytotherapy | Use of plant-based compounds (e.g., garlic, curcumin, neem extract) with antimicrobial properties to control infections | Promising results, especially in reducing infection rates of *S. agalactiae* without resistance risk | Variability in effectiveness depending on plant source, concentration, and application method; regulatory challenges | Generally cost-effective; lower costs than antibiotics; can be produced locally | Sustainable; reduces reliance on antibiotics and risk of resistance | Standardize dosages; use as part of an integrated management approach | Use of garlic extract and neem leaf extract in tilapia to reduce *S. iniae* incidence | 26,118-121 |
| Probiotics | Administration of beneficial bacteria to enhance the immune system and outcompete pathogens | Effective in improving overall fish health and reducing susceptibility to infections, especially for *S. iniae* and *S. agalactiae* | Requires careful selection of strains; inconsistent results across environments; regulatory approval needed | Moderate cost; can reduce the need for antibiotics, providing long-term savings | Sustainable; enhances fish health, potentially reducing the overall disease burden | Select strains based on specific farm conditions; monitor for efficacy regularly | Use of *Bacillus subtilis* and *Saccharomyces cerevisiae* in tilapia farming to prevent *Streptococcus* infections | 3,117,122-124 |
| Environmental Management | Management of water quality (e.g., temperature, pH, salinity) to reduce stress and prevent outbreaks | Crucial in preventing stress-induced susceptibility to infections, particularly *S. ictaluri* | Requires continuous monitoring and adjustment; influenced by external factors like | Initial costs for monitoring equipment; cost-effective in reducing disease outbreaks and mortality | Sustainable; essential for long-term aquaculture health; mitigates effects of climate change | Regular monitoring; adapt management practices to seasonal variations and climate trends | Managing water temperature in tilapia farms reduced *S. agalactiae* outbreaks | 47,125 |
| | | | climate change | | | | | |
| Immune Stimulants | Use of compounds that boost the fish immune system to enhance resistance to infections | Shows potential in increasing resistance to infections without the drawbacks of antibiotics | Limited studies on long-term efficacy; potential regulatory hurdles; variability in response | Initial investment in research and product development; long-term savings from reduced disease outbreaks | Potentially sustainable; reduces reliance on antibiotics and other chemical treatments | Integrate with other management strategies for best results; monitor for long-term effects | Use of β-glucans as immune stimulants in salmon farming to enhance resistance to Streptococcus infections | 126,127 |
| Genetically Resistant Strains | Breeding or genetically engineering fish strains that are resistant to Streptococcus infections | Highly effective in reducing disease incidence, particularly in species like tilapia and salmon | High initial research and development costs; ethical and regulatory concerns; potential ecological impacts | High upfront cost but offers long-term savings by reducing disease-related losses | Sustainable if managed responsibly; could revolutionize disease management in aquaculture | Combine with biosecurity and environmental management for comprehensive protection | Development of Streptococcus-resistant tilapia strains in Southeast Asia | 33,70,128 |
| Quorum Sensing Inhibitors (QSIs) | Disruption of bacterial communication to prevent coordination of virulence factors and biofilm formation | Promising in reducing virulence and biofilm-related infections | Early-stage research; more studies needed to establish efficacy in diverse aquaculture settings | High initial R&D costs; long-term cost-effectiveness depends on successful application | Could reduce reliance on antibiotics by targeting virulence without inducing resistance | Combine with other strategies; monitor for long-term impacts | Investigated as an alternative approach to reduce Streptococcus virulence | 129,130 |
| Phage Therapy | Use of bacteriophages to target and kill specific Streptococcus pathogens | High specificity; effective in controlling infections without harming beneficial microbiota | Still in experimental stages; challenges in phage selection, stability, and delivery | Potentially cost-effective with mass production; low environmental impact | Could offer a sustainable alternative to antibiotics; reduced risk of resistance | Requires development of phage libraries; integrate with biosecurity measures | Investigated as an alternative therapy to manage Streptococcus infections in aquaculture | 131,132 |
| Heat Shock Proteins (HSPs) | Induction of HSPs to enhance cellular protection and immune response | Potential in increasing resistance to infections, including Streptococcus | Practical application challenges; limited studies on aquaculture-specific use | Low cost if effective methods for induction are established; long-term benefits | Could improve overall fish health and resilience; potentially sustainable | Manage environmental conditions to induce HSPs without causing stress | Induction of HSPs in zebrafish increased resistance to Streptococcus infections | 133,134 |
| Monoclonal Antibody Therapy | Use of monoclonal antibodies to specifically target and neutralize Streptococcus pathogens | High specificity; can effectively neutralize pathogens without affecting beneficial microbiota | High production costs; regulatory challenges; requires precise targeting | High initial cost but could be justified in high-value operations | Could significantly reduce reliance on antibiotics; potential for long-term disease control | Requires development of specific antibodies; integrate with other management strategies | Use of monoclonal antibodies against Streptococcus iniae in rainbow trout | 135-137 |

Vaccination has emerged as a critical strategy to mitigate the significant challenges posed by *Streptococcus* infections in aquaculture, particularly in species like Nile tilapia and rainbow trout, which are highly susceptible to these bacterial pathogens. Various vaccination methods, including oral vaccines, live attenuated vaccines, and DNA vaccines, have been developed and tested for their effectiveness in increasing resistance and providing long-term protection against *Streptococcus* infections. A comparative overview of vaccination strategies, their effectiveness, cost-benefit considerations, and challenges is summarized in Table 6, detailing the various approaches used to prevent *Streptococcus* infections in fish.

Oral vaccines, especially those utilizing innovative coatings such as chitosan-alginate, have demonstrated promising results by enhancing immunity and improving survival rates. For instance, chitosan-alginate coated vaccines have been effective against *Lactococcus garvieae* and *S. iniae* in rainbow trout, with survival rates reaching 83% against *S. iniae* and 72% against *L. garvieae* [138]. Similarly, Eudragit L30D-55 encapsulated vaccines have provided high efficacy, with notable survival rates in treated fish. Feed-based vaccines are particularly advantageous due to their ease of administration and potential for mass vaccination. These vaccines have been effective in hybrid red tilapia, with bivalent vaccines incorporating inactivated *S. iniae* and *Aeromonas hydrophila* inducing strong immunological responses and high survival rates [139].

Injectable vaccines are a vital component of aquaculture vaccination programs, as they induce strong adaptive immune responses. Polyvalent injectable vaccines provide high survival rates and prolonged protection in Nile tilapia against multiple pathogens [140]. Live attenuated vaccines, such as those derived from *S. agalactiae*, have also demonstrated high relative percentage survival (RPS) values, reaching up to 90.47% in tilapia [141].

The effectiveness of vaccination strategies in aquaculture, including their long-term impact, cost-benefit considerations, and implementation guidelines, is further outlined in Table 5, which provides a broader overview of management strategies for *Streptococcus* infections, including vaccination as a key preventive measure.

Combining probiotics with vaccines has shown promise in enhancing immune responses in aquaculture. Probiotics such as *Bacillus subtilis* and *Lactobacillus plantarum*, when used alongside vaccination, significantly improved survival rates in Nile tilapia, reaching 97% against *S. agalactiae* [142]. Additionally, DNA vaccines represent a novel approach, inducing strong immune responses through the expression of specific antigens. An α-enolase-based DNA vaccine in Nile tilapia demonstrated significant protection, achieving a 72.5% survival rate against *S. iniae* and was associated with upregulation of immune-relevant genes [143].

Formalin-killed bacterin vaccines have also been successfully utilized in various fish species. In Asian seabass, such vaccines provided substantial protection, with no mortality reported in vaccinated fish and significant upregulation of innate immune genes post-challenge [144]. Recent advancements in vaccination strategies include the use of ozone nanobubbles (NB-O3) as a pre-treatment to enhance the efficacy of immersion vaccines, activating immune-related genes and improving survival rates in Nile tilapia [145]. Additionally, multi-epitope vaccines targeting various immunogenic proteins offer promising protection against *Streptococcus* infections [146].

Overall, vaccination strategies against *Streptococcus* infections in fish aquaculture are diverse and highly effective. Oral vaccines, feed-based vaccines, injectable vaccines, and innovative approaches such as DNA vaccines and probiotic-enhanced vaccination provide strong protection, enhancing fish health and reducing economic losses in aquaculture.

**Table 6.** Overview of vaccines and treatments used to manage *Streptococcus* infections in aquaculture, with details of their type, target species, mode of administration, and effectiveness. Challenges, availability, regulatory status, and implementation guidelines are addressed for each treatment. References to key studies support the effectiveness and practical application of these treatments.

| Vaccine/Treatment | Type | Target Streptococcus Species | Mode of Administration | Effectiveness | Challenges | Availability | Regulatory Status | Implementation Guidelines | References |
|---|---|---|---|---|---|---|---|---|---|
| AquaVac Strep Sa | Vaccine (Formalin-killed bacterin) | Streptococcus agalactiae | Intraperitoneal injection, immersion | High effectiveness (~70-90% reduction in mortality) in tilapia | Requires booster doses; handling stress during administration | Widely available in regions with high tilapia production | Approved in major markets (US, Asia, Latin America) | Administer early; avoid high-stress periods for fish | 11,108,141 |
| Florfenicol | Antibiotic | Broad-spectrum; used against Streptococcus iniae, S. agalactiae | Oral (medicated feed) | Effective when administered early; up to 80% recovery in trials | Resistance development; regulatory restrictions; withdrawal times | Widely available; approved in several countries | Approved in the US, EU, Asia; usage restrictions in some regions | Monitor for resistance; adhere to withdrawal periods | 147-149 |
| Oxytetracycline | Antibiotic | Broad-spectrum; used against various Streptococcus species | Oral (medicated feed) | Moderate effectiveness (~50-70% in trials); best when used early | Resistance development; environmental impact; regulatory issues | Widely available; increasingly restricted due to resistance concerns | Usage restricted in EU; still widely used in Asia and Latin America | Use as a last resort; monitor environmental impact | 27,147,150 |
| AquaVac Strep Si | Vaccine (Formalin-killed bacterin) | Streptococcus iniae | Intraperitoneal injection, immersion | High effectiveness (~75-85% reduction in mortality) in barramundi and tilapia | Requires booster doses; handling stress during administration | Widely available in Asia and Australia | Approved in major aquaculture markets | Administer during optimal temperature ranges to enhance efficacy | 108,140,151 |
| Phage Therapy | Biological Treatment | Streptococcus iniae, S. agalactiae | Immersion, oral | Promising in experimental studies; high specificity; | Experimental stage; challenges in phage | Limited availability; under development | Not yet approved; undergoing trials | Combine with biosecurity measures; ensure stable | 131,132 |

| | | | | | stability and delivery | | | phage formulation | |
|---|---|---|---|---|---|---|---|---|---|
| | | | | up to 90% reduction in bacterial load | | | | | |
| Ampicillin | Antibiotic | Broad-spectrum; effective against Streptococcus spp. | Oral, injection | Effective in early stages; less effective against resistant strains; ~60-75% recovery | Resistance issues; potential for environmental contamination | Available but use is declining due to resistance | Restricted in EU; available in US and Asia with caution | Use only when absolutely necessary; follow strict dosage guidelines | 36,112 |
| Garlic Extract | Phytotherapy | Streptococcus agalactiae | Oral (incorporated in feed) | Promising results in reducing infection rates; ~40-60% reduction in mortality | Variability in effectiveness depending on dosage and application | Commercially available as a feed additive | Generally approved as a feed supplement; varies by region | Standardize dosages; integrate with other treatments | 24,118,152 |
| β-glucans | Immune Stimulant | Broad-spectrum; boosts resistance to Streptococcus spp. | Oral, immersion | Effective as a preventive measure; boosts immune response; ~30-50% reduction in mortality | Variable efficacy; dependent on fish species and environmental factors | Available as an additive in some aquafeeds | Approved as a feed additive in most regions | Use as part of a broader health management strategy | 126,127 |
| Nanoparticle-Based Delivery Systems | Advanced Treatment | Streptococcus spp. | Injection, oral | Experimental; high potential for targeted delivery; enhanced drug stability | High R&D costs; regulatory hurdles; limited trials in aquaculture | Experimental; limited availability | Not yet approved; under research | Requires precise formulation; monitor for safety and efficacy | 153,154 |

## 4.2   Therapeutic interventions

### 4.2.1   Antibiotic treatments

In treating *Streptococcus* infections in aquaculture, several antibiotics have demonstrated significant efficacy. Amoxicillin, a widely used broad-spectrum penicillin, is highly effective against Gram-positive bacteria such as *Streptococcus* spp. This antibiotic has shown considerable success in reducing mortality rates in species like tilapia and catfish when administered properly. Its broad-spectrum activity ensures that it addresses multiple bacterial pathogens in aquaculture environments [111]. Erythromycin, a macrolide antibiotic, works by inhibiting bacterial protein synthesis and has been particularly effective against *Streptococcus* spp., significantly reducing bacterial loads and improving survival rates in species such as olive flounder and tilapia [112]. A comparative overview of antibiotic treatments, including their effectiveness, challenges, and cost-benefit considerations, is summarized in Table 5, which outlines various management strategies for *Streptococcus* infections. The table highlights commonly used antibiotics, their impact on fish health, and concerns regarding antibiotic resistance. While antibiotic treatments are effective, their long-term sustainability is questionable due to growing resistance issues and environmental concerns.

Gentamicin, an aminoglycoside antibiotic, is notable for its ability to treat severe bacterial infections. A combination of gentamicin with hypoionic shock has been explored to enhance antibiotic uptake by bacterial cells. This method has demonstrated rapid eradication of pathogens such as *S. iniae*, significantly improving health outcomes in infected fish [113,114]. Similarly, tylosin, another macrolide antibiotic, has been effectively used against *S. parauberis* in olive flounder. Both intramuscular and oral administration of tylosin have successfully reduced bacterial loads and enhanced survival rates, making it a valuable tool in managing streptococcal infections in aquaculture [155,156].

Oxytetracycline, a tetracycline antibiotic, is frequently used due to its broad-spectrum activity against *Streptococcus* spp. Its mechanism of action, inhibiting bacterial protein synthesis, makes it an effective choice for controlling outbreaks and maintaining fish health [150]. However, its overuse has been linked to increasing antibiotic resistance, leading to restricted usage in some regions. Florfenicol, a derivative of thiamphenicol, is favored for its strong antibacterial activity and minimal impact on human health through fish consumption. This antibiotic has been successfully used to treat a range of bacterial infections in aquaculture, including those caused by *Streptococcus* spp. [113]. The effectiveness of florfenicol in reducing *Streptococcus* infections is further outlined in Table 6, which compares various antibiotic treatments, their availability, and regulatory status in aquaculture settings.

Similarly, enrofloxacin, a fluoroquinolone antibiotic, has been employed to treat streptococcal infections in fish. It works by inhibiting bacterial DNA gyrase, which is essential for DNA replication and transcription. Enrofloxacin has shown high efficacy in controlling infections and improving survival rates across various fish species [60]. However, like oxytetracycline, enrofloxacin faces increasing regulatory restrictions due to resistance concerns.

The efficacy of antibiotic treatments compared to vaccination strategies is detailed in Table 6, which outlines various vaccines and treatments for *Streptococcus* infections in aquaculture. While antibiotics are effective for short-term control, vaccination provides a more sustainable long-term solution, reducing antibiotic dependence and minimizing the risk of resistance development. For instance, AquaVac Strep Sa, a formalin-killed bacterin vaccine, has shown 70-90% effectiveness in reducing mortality due to *S. agalactiae* in tilapia [11,107,108]. This emphasizes the need to shift from antibiotics to preventive measures such as vaccination, probiotics, and immune stimulants.

While these antibiotic treatments have proven effective, their application must be carefully managed to ensure sustained efficacy and minimize resistance risks. Antibiotic stewardship programs, including responsible usage, monitoring resistance patterns, and regulatory compliance, are crucial for ensuring the long-term viability of antibiotic treatments in aquaculture.

### 4.2.2   Alternative treatments (e.g., probiotics, herbal remedies)

The increasing prevalence of antibiotic-resistant *Streptococcus* strains has driven the exploration of alternative treatments, such as probiotics, herbal remedies, and other natural products. These alternative strategies have gained traction due to their ability to enhance immunity, reduce bacterial loads, and minimize antibiotic dependence. A comparative overview of these alternative treatments, including their effectiveness, cost-benefit considerations, and challenges, is provided in Table 5, which summarizes various management strategies for *Streptococcus* infections.

Probiotics have shown significant potential in managing streptococcal infections in fish by balancing gut microbiota, enhancing immune responses, and inhibiting pathogenic bacteria through competitive exclusion [123]. For instance, probiotic mixtures containing *Bacillus subtilis*, *Saccharomyces cerevisiae*, and *Aspergillus oryzae* have been shown to significantly improve immune responses and reduce mortality rates in tilapia infected with *S. iniae* [122]. Similarly, administering probiotics, prebiotics, and synbiotics through feed has enhanced survival rates and reduced bacterial loads in the organs of tilapia infected with *S. agalactiae* [124].

A detailed analysis of probiotic-based treatments, their mode of action, and their regulatory status in aquaculture is outlined in Table 6, which compares vaccines and treatments for *Streptococcus* infections. Probiotics are favored as a sustainable solution due to their ability to stimulate the fish's immune system and outcompete pathogens, reducing the need for antibiotic treatments.

Similarly, herbal remedies are being explored for their antimicrobial, anti-inflammatory, and immunostimulatory properties in controlling streptococcal infections. Essential oils from oregano, thyme, and cinnamon exhibit strong antibacterial activity against *Streptococcus* species by disrupting bacterial membranes and inhibiting biofilm formation [157]. Garlic extract, rich in allicin, has significant antibacterial effects against *S. iniae* and *S. agalactiae* [118]. Curcumin (from turmeric) and neem extract (rich in azadirachtin) significantly reduce bacterial counts and improve survival rates while enhancing immune responses in species such as silver catfish and tilapia [119,120,158]. Other natural products, such as Aloe vera and *Salvia officinalis* extracts, have been reported to reduce mortality rates and boost immune parameters, including lysozyme and peroxidase activities in infected fish [26,121].

The effectiveness of phytotherapy and probiotics in controlling *Streptococcus* infections is further illustrated in Table 5, which highlights their mode of action, cost-effectiveness, and long-term sustainability. While antibiotics provide immediate treatment, these alternative strategies offer a long-term, environmentally friendly solution to enhance disease resistance and improve aquaculture sustainability. Additionally, the combination of probiotics and herbal remedies has demonstrated synergistic effects in improving disease resistance. Herbal hydrogels encapsulating probiotics such as *Enterococcus faecium* have shown antimicrobial activity against *S. iniae*, significantly improving resistance in red hybrid tilapia [159]. This synergistic approach leverages the immunostimulatory properties of probiotics and herbal compounds, making it a viable strategy for managing bacterial infections in aquaculture.

Integrating these alternative treatments into aquaculture practices provides a sustainable approach to managing *Streptococcus* infections, enhancing fish immunity and reducing antibiotic dependence. Further advancements in herbal and probiotic-based treatments could reshape disease management strategies, reinforcing biosecurity and disease prevention efforts in aquaculture.

## 4.3    Case studies of successful disease management

Several case studies highlight effective management strategies that have significantly improved fish health and reduced economic losses in aquaculture. Table 5 provides an overview of various management strategies, including biosecurity, vaccination, probiotics, and alternative treatments, illustrating their effectiveness and implementation in different aquaculture settings.

In Malaysia, a comprehensive biosecurity program combined with vaccination and probiotics significantly improved health outcomes in red hybrid tilapia. Probiotics such as *Bacillus subtilis*, *Saccharomyces cerevisiae*, and *Aspergillus oryzae* were added to feed, enhancing the immune response and lowering mortality rates. This approach, coupled with strict biosecurity measures, led to a notable reduction in *Streptococcus* infections and improved overall fish health [117]. In another study from the Philippines, integrating probiotics with regular monitoring and good husbandry practices demonstrated significant improvements in managing *Streptococcus* infections in Nile tilapia. *B. subtilis* and *Lactobacillus plantarum* were used, resulting in higher survival rates and enhanced immune responses, including elevated levels of IgM antibodies [3].

Antibiotic therapy remains a primary control measure for *Streptococcus* infections. In China, a large-scale tilapia farming operation utilized a combination of antibiotics for acute outbreaks, followed by the implementation of a vaccination program for long-term protection. Florfenicol was particularly effective in treating *S. agalactiae* infections, with dosages of 20 and 40 mg/kg helps in controlling mortality during the treatment period [47]. The integration of antibiotic therapy with vaccination, as outlined in Table 6, controlled the immediate outbreak and prevented future occurrences, ensuring sustainable fish production and economic stability.

Selective breeding for disease resistance offers a complementary approach to managing *Streptococcus* infections. A study on Nile tilapia demonstrated substantial genetic variation in resistance to *S. iniae*, suggesting that selective breeding could improve disease resistance and provide commercial fish farmers with more resilient stock [135]. This strategy has shown promise in enhancing overall fish health and productivity and is further detailed in Table 5, which outlines genetic resistance strategies in disease management.

Vaccination is a critical component of *Streptococcus* infection management in aquaculture. A notable example is the development of a live attenuated vaccine against *S. agalactiae* for farmed Nile tilapia. This vaccine, derived from a non-encapsulated *S. agalactiae* strain (Δcps), has proven to be safe and effective, with vaccinated tilapia exhibiting high antibody titers and a significant immune response, leading to a relative percent survival (RPS) of 90.47% against *S. agalactiae* challenges [110]. This high efficacy is comparable to commercial vaccines, such as AquaVac Strep Sa, which has reduced mortality rates by 70-90% in tilapia, as detailed in Table 6.

In Japan, commercial vaccines have been licensed to prevent infections caused by *S. iniae, S. parauberis, S. dysgalactiae,* and *L. garvieae* in various aquaculture species. These vaccines have been instrumental in reducing mortality rates in species such as yellowtail, red sea bream, and flounder [160]. The implementation of vaccination programs has significantly improved the health and productivity of aquaculture operations, reinforcing the importance of vaccination in disease control. Similarly, a polyvalent vaccine tested on Nile tilapia broodstock and offspring, which included multiple pathogens such as *S. agalactiae*, improved non-specific and adaptive immunity, resulting in increased leukocyte counts, phagocytosis, lysozyme activity, and antibody titers. Immersion vaccination of larvae provided protection for up to three months, with an RPS of at least 60% [109].

Alternative therapies, including medicinal herbs and bacteriophages, have shown potential in managing *Streptococcus* infections. In Egypt, the incorporation of neem leaf extract and *Aloe vera* into fish diets significantly reduced bacterial load and improved survival rates in Nile tilapia infected with *S. iniae*. These herbal remedies enhanced the fish's immune response and reduced mortality, demonstrating the potential of natural treatments in bacterial infection control [121,161]. In Thailand, a fish farm combined vaccination with herbal supplements to manage *S. agalactiae* infections in hybrid tilapia. Garlic and curcumin extracts were incorporated into the fish diet, enhancing immune function, leading to higher survival rates and reduced disease incidence [162]. The success of herbal-based treatments in disease management, including their effectiveness and sustainability, is further detailed in Table 5, which outlines phytotherapy approaches for *Streptococcus* control.

Each disease management strategy presented in these case studies offers unique benefits, and when combined, they provide a comprehensive solution to the persistent challenge of *Streptococcus* infections in aquaculture. Continuous research and adaptation of these strategies to local conditions are essential for maintaining successful aquaculture operations and minimizing the impact of infectious diseases.

# 5.0  Challenges in disease management

## 5.1  Antibiotic resistance

### 5.1.1  Mechanisms of resistance

Antibiotic resistance in *Streptococcus* infections in fish is a growing concern in aquaculture. The mechanisms behind this resistance are complex, painting a challenging picture for researchers and aquaculture practitioners. At the heart of antibiotic resistance lies the ability of bacteria to alter their genetic makeup, often through horizontal gene transfer. This process, akin to bacterial espionage, involves the transfer of resistance genes from one bacterium to another, often facilitated by mobile genetic elements such as plasmids, transposons, and integrons. These elements can carry multiple resistance genes, enabling bacteria to rapidly acquire resistance to various antibiotics. For instance, genes like ermB, conferring resistance to macrolides, and tetM, responsible for tetracycline resistance, have been identified in *S. agalactiae* and *S. iniae* isolates from fish [163].

Mutations in chromosomal genes also play an important role in antibiotic resistance. These mutations can alter the target sites of antibiotics, rendering them ineffective. *Streptococcus* species, for instance, can modify the structure of penicillin-binding proteins (PBPs), reducing the affinity of beta-lactam antibiotics such as penicillin and amoxicillin. Similarly, mutations in DNA gyrase (gyrA) and topoisomerase IV (parC) genes confer resistance to quinolones, diminishing the efficacy of these drugs [110]. Efflux pumps further complicate the battle against antibiotic resistance. These membrane proteins act as bacterial bouncers, expelling antibiotics from the cell and reducing their intracellular concentration to sub-lethal levels. Efflux pumps can confer resistance to a broad spectrum of antibiotics, including tetracyclines and macrolides, and are a significant hurdle in maintaining effective antibiotic therapy [164-166].

Biofilm formation adds another layer of complexity to antibiotic resistance. *Streptococcus* species are adept at forming biofilms on various surfaces within aquaculture environments, such as fish gills, skin, and equipment. These biofilms act as fortresses, protecting the bacteria from antibiotics and the host immune system. The extracellular matrix of the biofilm impedes the penetration of antibiotics, while the close proximity of bacterial cells within the biofilm facilitates the transfer of resistance genes [167]. Florfenicol resistance in *S. agalactiae* offers a striking example of the complex nature of antibiotic resistance. Research has shown that florfenicol resistance is closely related to the reduction of intracellular drug accumulation caused by ATP-binding cassette (ABC) transporters. These transporters effectively reduce the intracellular concentration of florfenicol, allowing the bacteria to survive and proliferate despite the presence of the antibiotic [71].

As the resistance mechanisms unfold, the picture becomes increasingly intricate. Efflux pumps, genetic mutations, horizontal gene transfer, and biofilm formation work in concert, creating a formidable barrier against antibiotic efficacy. The stakes are high, as the continued use of suboptimal antibiotic treatments can worsen resistance, making it ever more challenging to manage streptococcal infections in aquaculture.

### 5.1.2   Impact on treatment efficacy

Antibiotic resistance in *Streptococcus* impacts the efficacy of treatments for infections in aquaculture. Studies have highlighted that antibiotic resistance mechanisms in *Streptococcus* significantly impair the efficacy of commonly used antibiotics like tetracycline and erythromycin [98,112,163,168,169]. For instance, the presence of resistance genes such as tet(M) and erm(B) in *Streptococcus* isolates from fish has rendered these antibiotics less effective. This resistance leads to a notable increase in the minimum inhibitory concentration (MIC) required to achieve therapeutic effects. Even a slight rise in the MIC of *S. agalactiae* to oxytetracycline can drastically reduce the probability of curing the infection, leading to higher mortality and carrier rates. Research has shown that for *S. agalactiae* strains with MICs greater than 0.06 μg/mL, the chances of achieving a bacterial cure are significantly diminished, while the probability of fish becoming carriers increases dramatically [2].

Moreover, the use of antibiotics such as florfenicol, norfloxacin, and oxytetracycline in treating *Streptococcus* infections in Nile tilapia has demonstrated varying degrees of success. While florfenicol and norfloxacin have resulted in no mortalities in infected fish, oxytetracycline treatment still led to a 20% mortality rate. This variability highlights the challenge posed by antibiotic resistance, as different *Streptococcus* strains exhibit varying sensitivity to these antibiotics [170]. The impact of antibiotic resistance extends beyond immediate treatment failures. Subclinical concentrations of antibiotics, such as streptomycin, disrupt the normal microbiome of fish, reducing microbial diversity and increasing early mortality rates. This disruption not only compromises the health of the fish but also creates reservoirs of resistance genes. Exposure to streptomycin has been shown to increase the abundance of class 1 integrons, genetic elements that facilitate the horizontal transfer of resistance genes, further complicating treatment efforts [171].

The economic implications of antibiotic resistance are substantial. Increased use of antibiotics to combat resistant strains leads to higher costs for fish farmers, both in terms of medication and the indirect costs associated with reduced fish growth and prolonged recovery times. Additionally, the environmental release of antibiotics can disrupt microbial communities and promote the development of resistance in non-target bacteria, posing long-term ecological risks [71]. The climax of this narrative lies in the urgent need for innovative strategies to address the profound impact of antibiotic resistance on treatment efficacy. Reducing antibiotic use, implementing proper dosing strategies, and exploring alternative treatments such as vaccines and probiotics are crucial but not definitive steps. Continued research and development of innovative treatment and prevention approaches are essential to mitigate the impact of resistance and ensure the sustainability of aquaculture practices [73].

## 5.2    Environmental and operational challenges

### 5.2.1    Water quality management

Water quality management is crucial in controlling *Streptococcus* infections in aquaculture. Parameters such as temperature, dissolved oxygen, and pH significantly impact the presence and severity of *Streptococcus* infections, as summarized in Table 7, which details the effects of various environmental factors on infection severity, prevalence, and long-term fish health.

Unfavorable conditions, such as high levels of organic matter and low dissolved oxygen, promote bacterial growth and disease outbreaks. Amal et al. [21] found that such conditions increase *S. agalactiae* presence in cultured tilapia, highlighting the importance of maintaining optimal water quality to prevent disease outbreaks. Temperature fluctuations also pose challenges; sudden changes or prolonged exposure to suboptimal temperatures weaken the fish's immune system, making them more susceptible to infections. As shown in Table 7, higher water temperatures (>28°C) are associated with increased infection rates, while optimal temperatures for fish health range between 22-28°C [11,21]. Maintaining stable water temperatures within this range is essential to prevent thermal stress and related health issues [21,23].

The effectiveness of disinfectants is another significant challenge under varying water quality conditions. Mon-On et al. [172] found that disinfectants such as povidone iodine, quaternary ammonium compounds, and glutaraldehyde are less effective in environments with higher organic matter and lower temperatures. Accumulated organic matter and suspended solids further complicate water quality management by creating breeding grounds for *Streptococcus* species. As outlined in Table 7, excessive organic loads increase pathogen proliferation, requiring effective filtration and sedimentation processes to remove excess organic matter. Overfeeding and inadequate waste management exacerbate this issue, stressing the need for stringent practices to manage waste and organic load in aquaculture systems [173].

Moreover, the presence of multidrug-resistant bacteria in water sources poses a serious challenge to effective disease management. Studies have documented the prevalence of antibiotic-resistant *Streptococcus* strains in various aquatic environments, complicating treatment efforts and increasing health risks [163]. As indicated in Table 7, high ammonia levels (>0.5 mg/L) cause stress and damage to fish gills, further increasing susceptibility to bacterial infections [174]. Implementing proper biofiltration and regular water exchanges is crucial for reducing ammonia accumulation and maintaining fish health.

Other key water quality factors influencing *Streptococcus* infections include dissolved oxygen, pH levels, and salinity. Low dissolved oxygen levels (<5 mg/L) increase fish stress and susceptibility to infections, whereas higher oxygen levels help reduce disease severity [21,175]. Additionally, pH fluctuations can directly affect pathogen survival and fish immunity, with optimal pH ranges being essential to prevent outbreaks [176]. As illustrated in Table 7, higher salinity (>20 ppt) inhibits *S. iniae* growth, whereas low salinity environments (<10 ppt) increase infection rates, making salinity adjustments an effective disease control measure [21,23,177].

Managing water quality is complex but essential for controlling *Streptococcus* infections in aquaculture. Advanced water quality management strategies, such as automated pH control systems, real-time oxygen monitoring, and improved waste filtration, are critical for sustaining fish health and preventing disease outbreaks. As outlined in Table 7, implementing proper environmental management strategies, including aeration, waste management, and regular water testing, is key to reducing *Streptococcus* infections and improving long-term aquaculture sustainability.

**Table 7.** Impact of environmental factors on *Streptococcus* species in aquaculture, with details of their effects on infection severity and prevalence. Examples of specific conditions and their long-term impacts on fish health are provided, along with mitigation strategies and implementation guidelines to manage these environmental challenges effectively. References to relevant studies are included to support the findings and recommendations presented.

| Environmental Factor | Impact on Streptococcus Species | Effect on Infection Severity/Prevalence | Examples | Long-Term Impacts | Mitigation Strategies | Implementation Guidelines | References |
|---|---|---|---|---|---|---|---|
| Water Temperature | S. iniae, S. agalactiae | Higher temperatures (>28°C) increase infection rates and severity; optimal temperatures for these species range between 22-28°C | High mortality in tilapia during summer months due to elevated temperatures | Persistent high temperatures can lead to chronic infection issues and reduced fish health | Adjust stocking density during peak temperature periods; use shading or cooling systems | Regular monitoring with temperature loggers; adjust feeding rates to reduce stress during high temperatures | 21,45 |
| pH Levels | Streptococcus spp. (general) | Low pH (<6.5) can increase stress and susceptibility to infection; high pH (>8.5) may reduce bacterial growth but can also stress fish | Increased Streptococcus infections in tilapia farms with fluctuating pH levels | Long-term exposure to suboptimal pH can weaken immune response and increase disease susceptibility | Regularly monitor pH and adjust with buffering agents; avoid rapid pH changes | Implement automated pH control systems in intensive aquaculture settings | 21,23,176 |
| Salinity | S. iniae, S. agalactiae | Lower salinity levels (<10 ppt) are associated with higher infection rates; high salinity (>20 ppt) can inhibit S. iniae growth | Outbreaks in freshwater systems, reduced prevalence in brackish waters | Long-term salinity management can help control pathogen prevalence in brackish systems | Gradual acclimation of fish to higher salinity; monitor for osmoregulatory stress | Regular salinity checks; use brackish water to manage Streptococcus outbreaks | 45,177 |
| Dissolved Oxygen | Streptococcus spp. (general) | Low dissolved oxygen (<5 mg/L) increases stress and susceptibility to infections; higher oxygen levels can reduce disease severity | Hypoxic conditions linked to higher mortality rates in catfish | Chronic low oxygen can lead to increased susceptibility to multiple pathogens, not just Streptococcus | Increase aeration and water flow; reduce feeding during low oxygen events | Install and maintain aeration systems; monitor oxygen levels continuously | 45,175 |
| Stocking Density | S. agalactiae, S. iniae | High stocking densities increase stress, reduce water quality, and elevate infection rates; optimal | Increased outbreaks in high-density | Long-term high-density farming without adequate management | Optimize stocking density according to species and | Regular density assessments; use biofilters to | 60,178 |

| | | | | | | |
|---|---|---|---|---|---|---|
| | | stocking density depends on species and system | tilapia and trout farms | increases chronic disease risk | environmental conditions | maintain water quality | |
| Water Flow and Circulation | Streptococcus spp. (general) | Poor water circulation can lead to localized hypoxic zones and higher bacterial loads, increasing infection risks | Stagnant areas in ponds associated with higher disease incidence | Long-term poor circulation can lead to chronic low-level infections | Improve water flow through pond design; use paddlewheels or pumps | Design systems with adequate circulation; regularly clean and maintain water channels | 179,180 |
| Organic Load/Nutrient Levels | Streptococcus spp. (general) | High organic load and nutrient levels can promote bacterial growth and reduce water quality, exacerbating infections | Fish farms with high organic loads reported more frequent infections | Persistent high organic loads can lead to eutrophication, increasing the prevalence of infections | Implement waste management practices; reduce feeding rates to minimize waste | Use of biofilters and regular sediment removal; monitor nutrient levels | 179,180 |
| Ammonia Levels | Streptococcus spp. (general) | High ammonia levels (>0.5 mg/L) cause stress and damage gills, increasing susceptibility to infections | High ammonia levels linked to increased disease outbreaks in intensive systems | Long-term exposure to ammonia can weaken fish health and reduce growth rates | Maintain proper biofiltration; regular water exchanges to reduce ammonia | Install ammonia monitoring systems; adjust feeding to control ammonia production | 21,23 |
| Light Intensity | Streptococcus spp. (general) | High or fluctuating light intensity can increase stress and weaken the immune response, increasing infection rates | Increased stress and disease prevalence in fish exposed to high light intensity | Managing light exposure can help maintain fish immune function and reduce stress | Use shaded areas or reduce light intensity in high-exposure environments | Monitor light levels in fish tanks or ponds; adjust artificial lighting as needed | 181,182 |

## 5.2.2 Farming practices and their impact on disease control

Farming practices in aquaculture play a salient role in disease control, and influences fish health, water quality, and pathogen prevalence. Several factors such as stocking density, feeding strategies, biosecurity protocols, and system design directly impact the incidence of *Streptococcus* infections. Table 7 provides an overview of how farming practices contribute to disease outbreaks, and highlights key environmental stressors and their long-term effects on aquaculture health.

One primary consideration is stocking density. High densities lead to overcrowding, increased fish stress, and suppressed immune function, making fish more vulnerable to infections. Studies indicate that tilapia stocked at higher densities exhibit higher mortality rates from *Streptococcus* infections compared to those kept at optimal stocking levels [178]. As shown in Table 7, high stocking densities elevate stress, leading to higher pathogen loads and reduced immune defenses. Optimizing stocking density based on species requirements and environmental conditions is crucial for reducing the risk of bacterial infections [60,135].

Feeding practices also play a vital role in disease control. Overfeeding increases organic matter accumulation, degrading water quality and creating favorable conditions for pathogenic bacteria. Excess nutrients promote eutrophication, as noted in Table 7, which exacerbates bacterial infections and disrupts aquatic ecosystems. Additionally, the nutritional content of feed influences fish immunity; balanced diets enriched with immunostimulants and probiotics have been shown to enhance resistance to bacterial infections [183,184]. Research demonstrates that incorporating probiotics such as *Lactobacillus* and *Bacillus* species into feed significantly reduces bacterial disease incidence and improves fish survival rates [183].

Biosecurity measures are another critical aspect of farming practices. Implementing strict biosecurity protocols, such as disinfecting equipment, proper fish handling, and quarantining new stock, prevents the introduction and spread of pathogens [185]. As outlined in Table 7, inadequate biosecurity and poor farming hygiene significantly increase pathogen transmission, emphasizing the need for comprehensive farm management strategies.

su

Water exchange practices are also integral to disease management. Regular water exchanges maintain optimal water quality by diluting waste products and reducing harmful pathogen concentrations. Farms that follow scheduled water exchange programs experience lower incidences of bacterial infections [186]. Table 7 details how poor water flow and stagnant areas create localized hypoxic zones, which increase bacterial proliferation, leading to higher mortality rates.

The design and maintenance of aquaculture systems further impact disease control. Well-designed systems that facilitate efficient water circulation, reduce organic load, and allow easy maintenance significantly lower bacterial loads. Regular tank and pond maintenance prevents bacterial buildup, reducing the risk of *Streptococcus* infections [187]. Table 7 highlights how inadequate circulation exacerbates pathogen survival, making proper system design a critical component of disease management.

Antimicrobial use is another important but often overlooked aspect of aquaculture. Misuse of antimicrobials leads to the emergence of antimicrobial-resistant bacteria, posing risks to both human and animal health [188]. The overuse of antibiotics is noted in Table 7 as a driving factor for antibiotic resistance, requiring alternative strategies such as probiotics and immunostimulants. Probiotics such as *Lactobacillus* and *Bacillus* species have been shown to enhance fish immunity, improve water quality, and promote growth, reducing the need for antibiotics in disease management [122].

Effectively maintaining farming practices that align with optimal environmental conditions enhances fish health and productivity, contributing to sustainable aquaculture operations. As indicated in Table 7, implementing best practices in stocking density, feeding, biosecurity, water quality, and system maintenance is essential for reducing *Streptococcus* infections and improving overall fish health.

## 6.0  Future directions in research and management

### 6.1  Innovative approaches to prevention and treatment
#### 6.1.1  Genetic engineering and selective breeding
Genetic engineering and selective breeding represent groundbreaking strategies for enhancing disease resistance in aquaculture species, particularly against *Streptococcus* infections. Selective breeding has long been instrumental in improving desirable traits such as growth rates, feed efficiency, and disease resistance in farmed fish. By selectively breeding individuals with natural resistance to streptococcal infections, aquaculture operations can develop populations with enhanced resilience [160]. For instance, selective breeding programs for Nile tilapia have successfully produced strains with increased resistance to *S. agalactiae*, leading to lower mortality rates and improved overall health. In such programs, selected lines demonstrated a 65% reduction in mortality risk compared to unselected lines, emphasizing the potential of genetic selection in disease resistance. These methodological approaches used in *Streptococcus* research, particularly genetic selection and breeding programs, are detailed in Table 8, which outlines various research techniques in aquaculture genetics.

Additionally, selective breeding for resistance to *S. iniae* has shown promising results. Studies reveal substantial genetic variation in resistance to *S. iniae*, with heritability estimates as high as 0.58, suggesting strong potential for genetic improvement through selective breeding (Rutten et al., 2005). Marker-assisted selection (MAS) using significant SNPs linked to resistance traits has produced offspring with dramatically lower mortality rates, demonstrating the efficacy of this approach [160]. Genomic selection (GS) and genome-wide association studies (GWAS) have further enhanced precision breeding programs, identifying key quantitative trait loci (QTLs) associated with resistance to *S. iniae* and *S. agalactiae* [83].

Genetic engineering extends these advancements by enabling precise modifications to the fish genome. CRISPR-Cas9, a cutting-edge gene-editing tool, allows for direct modifications of genes associated with disease resistance, enhancing immunity to streptococcal infections. Recent studies have demonstrated the potential of CRISPR-Cas9 to create gene knockouts in zebrafish, significantly increasing resistance to bacterial pathogens [14]. However, the application of genetic engineering for resistance to *Streptococcus* infections in commercially farmed species such as tilapia and catfish remains an emerging research area. As highlighted in Table 8, CRISPR-Cas9 gene editing has been successfully applied to functional studies of *Streptococcus* virulence factors, providing insights into host-pathogen interactions and immune response mechanisms.

The integration of omics technologies, such as genomics, transcriptomics, and proteomics, further enhances genetic engineering and selective breeding strategies. These approaches enable comprehensive analysis of genetic and molecular profiles, identifying biomarkers and key pathways involved in disease resistance. For instance, transcriptomic analyses of resistant and susceptible tilapia populations have revealed differentially expressed immune-related genes, providing valuable targets for selective breeding and genetic modification [83]. Advances in whole genome sequencing (WGS) have also allowed for the detailed characterization of *Streptococcus* strains, enabling strain-specific resistance breeding and vaccine development, as documented in Table 8.

Innovative approaches in genetic engineering, such as synthetic biology and gene drives, hold future potential for disease management in aquaculture. Synthetic biology could enable the design of new genetic circuits to boost immune responses, while gene drives could rapidly spread beneficial resistance traits through fish populations. However, these technologies also raise ethical and ecological concerns, necessitating rigorous research and regulatory oversight before large-scale application. Field trials and bioinformatics-based computational models, as outlined in Table 8, can help assess the feasibility and long-term effects of genetic modifications in aquaculture environments.

Currently, there is limited evidence of successful genetic engineering specifically targeting *Streptococcus* resistance in fish species. While CRISPR-Cas9 has been used to enhance resistance to other pathogens, targeted genetic modifications for *Streptococcus* resistance remain underexplored. Future research should focus on identifying key immune genes associated with resistance, integrating genetic engineering with traditional breeding programs, and leveraging omics-based approaches for disease prevention. These methodological innovations, documented in Table 8, can provide a robust framework for developing disease-resistant fish populations, revolutionizing aquaculture health management

**Table 8.** Overview of various methodological approaches used in *Streptococcus* research within aquaculture, detailing their applications, strengths, limitations, and comparative insights. Each method is further contextualized with implementation guidance, examples of practical applications, and references to key studies. The methodologies range from molecular techniques and in vitro assays to field trials and advanced genomic tools, highlighting their contributions to understanding and managing *Streptococcus* infections in aquaculture.

| Methodological Approach | Study Type | Applications | Strengths | Limitations | Comparative Insights | Implementation Guidance | Examples | References |
|---|---|---|---|---|---|---|---|---|
| Molecular Techniques (PCR, qPCR, RT-PCR) | Experimental | Detection and quantification of Streptococcus DNA/RNA in fish tissues, water samples; gene expression analysis | High sensitivity and specificity; allows for quantification; can detect low levels of pathogen | Requires specialized equipment; potential for contamination leading to false positives | Highly effective for early detection; cost-effective in established labs | Ensure strict contamination control; use validated primers for Streptococcus species | Detection of S. iniae in farmed tilapia using qPCR | 12,39 |
| Whole Genome Sequencing (WGS) | Genomic | Characterization of Streptococcus strains; identification of virulence factors and resistance genes | Provides comprehensive genetic information; useful for tracking outbreaks | High cost; requires bioinformatics expertise; large data sets can be challenging to analyze | Best for strain typing and understanding genetic diversity; expensive but increasingly accessible | Use for detailed outbreak investigations; ensure proper data storage and bioinformatics support | Genomic analysis of S. agalactiae isolates from different aquaculture settings | 34,189 |
| Epidemiological Surveys | Field Study | Assessment of Streptococcus infection prevalence in aquaculture facilities; risk factor analysis | Provides real-world data; useful for large-scale monitoring and risk assessment | Can be time-consuming and resource-intensive; data can be influenced by confounding factors | Best for understanding large-scale patterns and identifying risk factors; requires careful design | Include adequate sample sizes and control for confounding variables; use standardized data collection tools | Prevalence survey of S. iniae in tilapia farms in Asia | 60,190 |
| In Vitro Assays (Antimicrobial Susceptibility Testing) | Laboratory | Testing the efficacy of antibiotics and other treatments against Streptococcus strains | Allows for controlled testing of treatment efficacy; can guide therapeutic decisions | In vitro conditions may not fully replicate in vivo environments; may not | Useful for preliminary testing of treatments; moderate cost and widely used | Use standardized protocols; validate results with in vivo models when possible | Susceptibility testing of S. agalactiae to various antibiotics | 191,192 |
| | | | | account for host factors | | | | |
| Experimental Infection Models | Experimental | Understanding the pathogenesis of Streptococcus infections; testing vaccine efficacy | Allows for controlled study of disease progression and treatment; can simulate natural infections | Ethical concerns; requires careful design to replicate natural infection conditions | Best for studying disease mechanisms and testing interventions; ethical considerations must be addressed | Ensure humane treatment of animals; use appropriate controls and replicate experiments | Experimental infection of tilapia with S. iniae to assess vaccine efficacy | 45,80 |
| Histopathology | Diagnostic/Analytical | Examining tissue samples to understand the pathology and tissue tropism of Streptococcus infections | Provides detailed information on tissue changes and disease progression | Requires specialized expertise; can be time-consuming | Highly informative for understanding disease impact at the tissue level; labor-intensive | Use standardized staining techniques; ensure proper sample fixation | Histopathological analysis of lesions in S. iniae-infected tilapia | 97,174 |
| Field Trials | Applied Research | Testing the effectiveness of vaccines, probiotics, or other treatments in real-world aquaculture settings | Provides practical insights into the effectiveness of treatments; accounts for environmental factors | Can be logistically challenging; results may be influenced by uncontrolled variables | Essential for validating treatment efficacy in commercial settings; requires careful planning | Include control groups; monitor environmental conditions closely; ensure replicability | Field trial of a S. agalactiae chevaccine in tilapia farms | 92,143,193 |
| Serological Methods (ELISA, Agglutination Tests) | Diagnostic | Detection of antibodies against Streptococcus species in fish serum; used for surveillance and monitoring | Non-lethal; allows for monitoring of immune responses over time | May not distinguish between current and past infections; requires | Useful for ongoing surveillance and immune response monitoring; requires validation | Use validated reagents and protocols; combine with other diagnostic methods for confirmation | ELISA-based detection of antibodies against S. iniae in barramundi | 54,194 |
| | | | | validated reagents | | | | |
| Bioinformatics and Computational Models | Analytical | Analyzing genomic data, predicting disease outbreaks, modeling disease dynamics | Can handle large datasets; useful for hypothesis generation and testing | Requires specialized software and expertise; results depend on the quality of input data | Increasingly important for large-scale data analysis and predictive modeling; requires expertise | Ensure high-quality input data; use validated models; collaborate with bioinformaticians | Modeling the spread of Streptococcus infections in aquaculture systems | 40,69,195 |
| CRISPR-Cas9 Gene Editing | Experimental /Genomic | Studying gene function in Streptococcus; potential for developing resistant strains | High precision; can target specific genes; useful for functional studies | Ethical and regulatory concerns; off-target effects; requires specialized equipment | Cutting-edge approach for genetic studies; expensive and technically demanding | Use carefully designed guide RNAs; ensure thorough validation of results | Functional analysis of virulence factors in S. agalactiae using CRISPR | 14,30 |
| Omics Approaches (Proteomics, Metabolomics) | Analytical/Ex perimental | Comprehensive analysis of proteins, metabolites in Streptococcus; understanding host-pathogen interactions | Provides detailed insights into cellular processes; can -- identify novel targets for treatment | Requires specialized equipment and expertise; data analysis can be complex | Ideal for discovering biomarkers and therapeutic targets; requires significant resources | Ensure proper sample preparation; collaborate with experts in data analysis | Proteomic analysis of S. iniae to identify virulence factors | 196,197 |

The development of vaccines in aquaculture, particularly against *Streptococcus* infections, has been propelled by recent advances in omics technologies. These innovations provide a deeper understanding of pathogen genomics and host immune responses, enabling the creation of more effective vaccines. High-throughput genomic analyses of pathogens like *Streptococcus suis* have revealed extensive genetic variability, which complicates vaccine design but also identifies critical antigen candidates with minimal sequence variation. These insights facilitate the development of vaccines tailored to diverse pathogen populations [198]. Table 8 highlights the methodological approaches used in vaccine development, including genome-wide studies, transcriptomics, and proteomics, which are essential for identifying novel vaccine targets and optimizing immune response strategies.

Traditional vaccines in aquaculture, often based on inactivated or attenuated pathogens, have limitations such as the requirement for multiple doses and limited duration of immunity. Recent advancements in subunit vaccines, which use specific pathogen-derived antigens, are addressing these challenges. Novel antigens from *S. agalactiae* and *S. iniae* have shown promising results, eliciting strong immune responses and reducing mortality rates in fish [33,72,151]. Recombinant protein vaccines, a type of subunit vaccine produced using recombinant DNA technology, have also demonstrated high levels of protection in vaccinated fish [86]. The efficacy of these emerging vaccine strategies is further outlined in Table 8, which details the use of in vitro and field trials in vaccine testing.

The advent of DNA vaccines marks a significant leap forward in fish vaccination. These vaccines involve the direct injection of genetic material encoding antigenic proteins, leading to in vivo production of these proteins and subsequent immune activation. DNA vaccines offer the advantage of inducing both humoral and cellular immunity and provide long-lasting protection. Studies have demonstrated that DNA vaccines significantly protect tilapia against *S. agalactiae*, highlighting their potential for widespread use in aquaculture [31]. As shown in Table 8, molecular techniques such as qPCR and transcriptomic profiling play an importrant role in evaluating vaccine efficacy, allowing researchers to assess immune response markers and pathogen load post-vaccination.

Advancements in adjuvant technologies have further enhanced vaccine efficacy. Novel adjuvants, such as nanoparticles and liposomes, improve antigen delivery and presentation, thereby boosting immune responses. Liposome-based adjuvants, for instance, have been shown to significantly enhance immunity against *S. iniae*, leading to better protection in fish [199]. Live attenuated vaccines, using genetically modified pathogens, offer another promising approach. Techniques such as targeted gene deletion have produced attenuated strains of *S. agalactiae* that provoke strong immune responses without causing disease, providing long-term protection with minimal side effects [14,30]. CRISPR-Cas9 gene editing, as described in Table 8, has been explored for developing genetically modified bacterial strains that induce protective immune responses while minimizing virulence.

Multi-omics approaches, integrating genomics, transcriptomics, proteomics, and metabolomics, are pivotal in identifying new vaccine targets and understanding immune mechanisms in fish. These comprehensive analyses facilitate the discovery of novel antigens and the development of more effective vaccines. For example, proteomic studies have identified key proteins involved in *S. iniae* pathogenicity, which are now being explored as vaccine candidates [47]. Whole genome sequencing (WGS), another key approach listed in Table 8, enables comparative genomic studies of vaccine-targeted bacterial strains, ensuring broad-spectrum protection against multiple pathogen serotypes.

Despite these advancements, challenges remain in optimizing vaccine development for practical use in aquaculture. Variability in immune responses among different fish species and the need for broad-spectrum protection against multiple pathogens present significant hurdles. The cost and logistics of vaccine administration in large-scale aquaculture operations also pose challenges. Table 8 outlines various epidemiological surveys and field trials, which are essential for evaluating vaccine effectiveness in real-world farming conditions. Future research should prioritize the exploration of novel antigens, refine delivery systems, and integrate vaccines with other disease management strategies. Computational models and bioinformatics approaches, as highlighted in Table 8, are becoming increasingly valuable in predicting vaccine efficacy and optimizing vaccine formulations before large-scale trials. The integration of advanced molecular tools and big data analytics into vaccine research promises further breakthroughs in the fight against *Streptococcus* infections in aquaculture.

## 6.2 Role of information technology in disease management

### 6.2.1 Big data analytics in fish health monitoring

Big Data Analytics has become an essential tool in aquaculture, particularly for disease monitoring and management. Real-time monitoring of water quality parameters plays a critical role in early disease detection and prevention. Internet of Things (IoT) devices and sensors continuously collect extensive data on key environmental factors such as temperature, pH, dissolved oxygen, and ammonia levels. This continuous data acquisition allows for the identification of trends, anomalies, and risk factors associated with *Streptococcus* outbreaks. As highlighted in Table 8, the use of bioinformatics and computational models helps analyze these large datasets, improving predictive capabilities for aquaculture disease management.

Big data frameworks, such as Hadoop and Hive, facilitate the storage and processing of vast datasets, enhancing real-time disease tracking. These tools help analyze historical trends to identify disease patterns, improving the accuracy and speed of disease diagnosis. For example, historical data analysis has effectively identified shrimp disease symptoms, demonstrating the practical applications of these frameworks in aquaculture [200]. Geographic Information Systems (GIS) further enhance disease monitoring by mapping and analyzing the spatial distribution of diseases. As outlined in Table 8, GIS-based reinforcement learning techniques such as the Multi-Armed Bandit (MAB) approach have been successfully used to predict disease transmission patterns in fish farms in Greece [201].

The integration of genomic data into disease surveillance has also transformed disease detection and prevention. High-throughput sequencing technologies and multi-omics approaches provide detailed insights into the genetic makeup of pathogens and host species. This genomic information is crucial for identifying genetic markers associated with disease resistance, allowing for precision breeding and improved vaccine development. As shown in Table 8, whole genome sequencing (WGS) is an indispensable tool for characterizing *Streptococcus* strains, enabling targeted intervention strategies and disease tracking [34]. Bioinformatics tools process this genomic data to identify novel vaccine candidates and therapeutic agents, accelerating the development of next-generation vaccines [40,69].

Despite advancements, challenges remain in applying big data analytics to aquaculture. Data integration and standardization issues, along with the need for advanced analytical skills, pose technical barriers. Additionally, ethical considerations related to data privacy, consent, and fairness must be addressed to ensure responsible data use in disease monitoring systems. Table 8 outlines the importance of computational modeling and machine learning, which can help overcome these limitations by automating data interpretation and prediction models.

Nevertheless, these efforts in big data analytics culminate in precise monitoring, early disease detection, and timely intervention for *Streptococcus* infections. The integration of advanced technologies, such as IoT sensors, GIS mapping, bioinformatics, and real-time predictive modeling, offers a promising future for managing streptococcal infections in aquaculture. As documented in Table 8, the synergy between big data analytics and disease management strategies is revolutionizing aquaculture health monitoring, enhancing the industry's sustainability and productivity.

### 6.2.2   Machine learning and predictive modeling

Machine learning and predictive modeling have revolutionized aquaculture by offering precise methods to anticipate and make better decisions when managing disease outbreaks. These technologies leverage vast datasets to identify patterns and make predictions that surpass traditional methods. As highlighted in Table 8, bioinformatics and computational modeling are critical methodologies for predicting disease transmission, optimizing treatment strategies, and improving disease management.

At the forefront, machine learning algorithms such as neural networks, support vector machines (SVM), and random forests (RF) analyze extensive aquaculture datasets. These algorithms learn from historical data and current environmental parameters, predicting disease outbreaks with high accuracy. For instance, predctive models can analyze water quality variables such as temperature and pH to forecast the likelihood of *Streptococcus* infections, enabling timely preventive measures [202]. Table 8 details the use of computational models in disease risk prediction, reinforcing the role of AI-driven methodologies in aquaculture disease monitoring.

One innovative application of machine learning is the early detection of *Streptococcus* infections in fish. By processing historical outbreak data and real-time environmental conditions, predictive models provide immediate alerts to farmers. This proactive approach allows for timely interventions, such as adjusting feeding practices or enhancing water quality, to mitigate infection risks. Studies demonstrate that machine learning models can predict *Streptococcus* outbreaks with high accuracy, thereby improving disease management strategies [203]. As documented in Table 8, epidemiological surveys and field trials are key tools in validating these predictive models, ensuring their practical application in commercial aquaculture.

Machine learning-based classification models have also transformed pathogen identification. For example, integrating MALDI-TOF MS with machine learning algorithms has proven effective in classifying bacterial serotypes. MALDI-TOF MS combined with SVM and RF algorithms can rapidly identify Group B *Streptococcus* (GBS) serotypes, achieving prediction accuracies between 54.9% and 87.1% [195]. Similarly, for *S. suis* serotype 2, machine learning-based classifiers have exceeded 90% accuracy, demonstrating their potential for high-speed, specific pathogen identification [204].

Beyond disease detection, predictive modeling also optimizes vaccination strategies. By simulating streptococcosis outbreaks, improvement during multiple vaccination scenarios, machine learning algorithms could help determine the optimal timing and frequency of vaccinations, thereby enhancing fish health and reducing antibiotic dependence. Additionally, machine learning models assist in identifying the most effective vaccine formulations by analyzing genetic and phenotypic data from diverse fish populations [205]. As shown in Table 8, computational modeling tools are increasingly used in vaccine development, particularly in identifying key antigenic targets and optimizing immunization protocols.

The integration of machine learning with IoT and big data analytics further amplifies its potential in real-time disease management. IoT devices continuously monitor water quality parameters, transmitting massive datasets to cloud platforms, where machine learning algorithms analyze trends in real time. This synergistic approach enables continuous monitoring and immediate responses to adverse environmental conditions, ensuring the health and sustainability of aquaculture systems. As detailed in Table 8, bioinformatics and computational tools are crucial in refining predictive models for practical disease control.

However, challenges persist in fully exploiting these technologies. Accurate predictions hinge on high-quality, extensive datasets, yet inconsistent data collection practices and a lack of standardization remain obstacles. Additionally, machine learning algorithms require specialized expertise, necessitating advanced training within the aquaculture industry [206]. Ultimately, machine learning and predictive modeling are transforming disease management in aquaculture by enabling early detection, optimizing vaccination schedules, and integrating seamlessly with IoT and big data analytics. As documented in Table 8, these technologies are at the forefront of modern aquaculture research, driving precision disease monitoring and sustainable disease prevention strategies.

## 6.3    Integrated disease management approaches

Integrated management of streptococcal infections in aquaculture necessitates combining the One-Health framework with effective collaboration among industry, academia, and government. The One-Health approach emphasizes the interconnectedness of human, animal, and environmental health, essential for tackling the complexities of infectious diseases in aquaculture. By recognizing that these sectors are interdependent, the One-Health approach promotes a holistic understanding of the disease dynamics affecting aquaculture. Environmental DNA (eDNA) surveillance, for example, can detect pathogenic DNA in water samples, providing early warnings for potential outbreaks. This proactive monitoring enables timely interventions to prevent the spread of infections [207,208].

Additionally, understanding the environmental conditions that favor *Streptococcus* species proliferation aids in designing better management practices to minimize infection risks. Factors such as water quality, temperature, and nutrient levels must be meticulously monitored and controlled to create an inhospitable environment for pathogens. The One-Health approach also highlights the human health implications of zoonotic diseases in aquaculture. Human health can be directly impacted by consuming infected fish or through occupational exposure among aquaculture workers. Therefore, improving fish health and maintaining environmental standards also protect public health, reducing the risk of zoonotic disease transmissions [16,209].

Collaboration among stakeholders is crucial for effective disease management. The aquaculture industry provides practical insights and resources essential for implementing disease control measures. Farmers play a critical role by adopting improved biosecurity practices, such as disinfecting equipment, managing water quality, and controlling stock densities to reduce stress and infection risks. Academia contributes through cutting-edge research and the development of innovative solutions. For instance, research institutions can conduct trials to test the efficacy of new vaccines and develop guidelines for their use in various aquaculture settings [86]. Government agencies are pivotal in regulating and supporting these efforts, facilitating the implementation of best practices, providing training programs for farmers, and ensuring compliance with biosecurity protocols. Governments can also fund research initiatives exploring new technologies for disease management and support infrastructure development for large-scale vaccination programs. Policies and regulations must evolve to keep pace with scientific advancements and industry needs, ensuring a supportive framework for integrated disease management [210,211].

Moreover, information technology plays a vital role in promoting the One-Health approach and fostering collaboration among stakeholders. Advanced technologies such as IoT, big data analytics, and machine learning enhance disease management capabilities by enabling real-time monitoring, early detection, and rapid response to disease outbreaks. For example, IoT devices can continuously monitor water quality parameters, while machine learning algorithms analyze data to predict disease outbreaks, optimizing resource allocation and decision-making processes. Integrating these technologies with the One-Health could enable stakeholder collaboration that ensures a more efficient and effective response to streptococcal infections [14]. This holistic strategy ensures that all aspects of the ecosystem are considered, promoting the health and sustainability of aquaculture systems.

## 7.0   Conclusion

This review highlights the critical need for a comprehensive, multifaceted approach to effectively manage and mitigate *Streptococcus* infections. Advancements in genetic engineering and selective breeding offer promising solutions for developing disease-resistant fish strains. Techniques such as CRISPR-Cas9 provide precise and efficient methods for enhancing genetic resistance to infections. Concurrently, the development of innovative vaccines, including subunit, DNA, and recombinant protein vaccines, demonstrates significant potential in providing targeted immunoprophylactic strategies tailored to specific pathogens. The integration of big data analytics and IoT technologies represents a transformative leap in disease monitoring and management. These tools enable real-time surveillance, predictive modeling, and timely interventions, significantly bolstering the resilience of aquaculture systems. Additionally, the One-Health approach, emphasizing the interconnectedness of human, animal, and environmental health, advocates for a holistic strategy in disease management, requiring strong collaboration among industry stakeholders, academic researchers, and government agencies to address infectious disease complexities comprehensively. Future research must focus on ongoing innovation and interdisciplinary collaboration to develop integrated management practices that combine technological advancements with sustainable farming techniques, ensuring the long-term sustainability and productivity of the aquaculture industry.

**Use of Generative-AI tools declaration**

The authors declare that they used ChatGPT (OpenAI) to improve the readability, clarity, and narrative flow of this manuscript. The AI tool was applied during the drafting and revision stages, specifically for language editing and paraphrasing suggestions. All scientitfic content, data, and interpretations were critically reviewed and edited by the authors, who take full responsibility for the scientific accuracy and integrity of the work. AI assistance was primarily used in the Abstract, Introduction, and Discussion sections.